\documentclass[a4paper]{book}

\usepackage{amsmath,amsthm,amssymb}
\usepackage{amssymb,amsmath,amsthm,graphicx,epsfig,latexsym}

\newtheorem{theorem}{Theorem}

\newtheorem{lemma}{Lemma}
\newtheorem{corollary}{Corollary}

\newtheorem{definition}{Definition}

\theoremstyle{remark}
\newtheorem{example}{Example}[section]

\newcommand\beq{\begin{equation}}
\newcommand\eeq{\end{equation}}
\newcommand\bea{\begin{eqnarray}}
\newcommand\eea {\end{eqnarray}}
\newcommand\ba{\begin{array}}
\newcommand\ea {\end{array}}

\newcommand\bd{\begin{description}}
\newcommand\ed {\end{description}}
\newcommand\ben{\begin{enumerate}}
\newcommand\een{\end{enumerate}}

\newcommand\bD{\begin{definition} }
\newcommand\eD{\end{definition} }
\newcommand\bE{\begin{example} }
\newcommand\eE{\end{example} }
\newcommand\bL{\begin{lemma} }
\newcommand\eL{\end{lemma} }
\newcommand\bT{\begin{theorem} }
\newcommand\eT{\end{theorem} }

\newcommand\bC{\begin{corollary} }
\newcommand\eC{\end{corollary} }

\def\a{{\alpha}}
\def\b{{\beta}}
\def\g{{\gamma}}

\def\th{{\theta}}
\def\r{{\rho} }

\def\G{{\Gamma} }

\def\d{{\partial} }

\def\v{{\nu} }
\def\n{{\textbf{n}} }
\def\mt{{m_t(r)} }
\def\mtp{{m_t(p)} }
\def\mv{{m_V} }

\def\g{{g_{toy}} }

\def\grr{{g_{rr}} }
\def\gtt{{g_{tt}} }

\def\grrp{{g_{rr}(p)} }
\def\gttp{{g_{tt}(p)} }

\def\Ktf{{K_{23}(r)} }
\def\Ktfp{{K_{23}(p)} }
\def\KS{{K_{S}(r)} }
\def\KSp{{K_{S}(p)} }

\def\ks{{K_{S}} }
\def\khf{{K_{23}} }
\def\krh{{K_{12}} }
\def\krt{{K_{01}} }
\def\kht{{K_{02}} }

\def\thd{{\dot{\theta}}  }

\def\ar{{\alpha(r)} }

\def\Sx{{\S_{\{X,x^0\}}} }
\def\Sxp{{\S_{\{X,x^0(p)\}}} }

\def\R{{\mathbb{R}}}
\def\Z{{\mathbb{Z}}}

\def\M{{\mathcal{M}}}
\def\TM{{T\M_{p}}}

\def\L{{\mathcal{L}}}
\def\S{{\mathcal{S} }}
\def\Ob{{\mathcal{O}}}

\def\nn{{\nonumber} }

\def\lap{{\triangle} }

\def\dGttr{{\nabla_1 G_{00}}  }

\def\Rtt{{R_{00}}  }

\newcommand{\ord}[1]{O({m}^{{#1}})}
\newcommand{\ordt}[1]{O({\mt}^{{#1}})}
\newcommand{\ordx}[1]{O({m_X}^{{#1}})}
\newcommand{\ordc}[1]{O({m_C}^{{#1}})}
\newcommand{\orda}[1]{O({m_A}^{{#1}})}

\newcommand{\ordv}[1]{O({m_V}^{{#1}})}

\begin{document}

\pagestyle{empty}
\begin{titlepage}

\newcommand*{\titleTMB}{\begingroup
\centering
\settowidth{\unitlength}{\LARGE QUANTUM MEASURE THEORY:}
\vspace*{\baselineskip}
\rule{\unitlength}{1.6pt}\vspace*{-\baselineskip}\vspace*{2pt}
\rule{\unitlength}{0.4pt}\\[\baselineskip]
{\LARGE Matter:}\\[\baselineskip]
{\LARGE Space Without Time}\\[\baselineskip]
\rule{\unitlength}{0.4pt}\vspace*{-\baselineskip}\vspace{3.2pt}
\rule{\unitlength}{1.6pt}\\[\baselineskip]\vspace{3.2pt}
{\Large\itshape Yousef Ghazi-Tabatabai \\ \vspace{3.2pt} yousef.ghazi05@imperial.ac.uk}\par
\par
\endgroup}

\titleTMB


\end{titlepage} 
\pagestyle{empty}\pagenumbering{roman}

\tableofcontents
\chapter[Matter: Space without Time]{Matter: Space without Time}\label{ch:MASWT}
\setcounter{page}{1}
\pagestyle{headings}
\pagenumbering{arabic}
\vskip 1cm

\begin{small}
\begin{center}
   \textbf{Abstract}
  \end{center}
While Quantum Gravity remains elusive and Quantum Field Theory retains the interpretational difficulties of Quantum Mechanics, we introduce an alternate approach to the unification of particles, fields, space and time, suggesting that the concept of matter as space without time provides a framework which unifies matter with spacetime and in which we anticipate the development of complete theories (ideally a single unified theory) describing observed `particles, charges, fields and forces' solely with the geometry of our matter-space-time universe.
\end{small}
\vskip 0.5cm

\section{Introduction}\label{sec:introduction}

\subsection{Through the Prism of Unification}\label{subsec:through the prism of unification}

Twentieth century physics was dominated by two `great paradigms', the intuitively elegant General Relativity (GR) and the intuitively confused Quantum Mechanics (QM) which developed into Quantum Field Theory (QFT). Both are advances upon what we will call the two core paradigms of the `classical' physics current at the end of the nineteenth century; the `snooker balls on a table' paradigm of discrete particles (central in Newtonian mechanics) and the `ripples on a pond' paradigm of continuous fields (prominent, for example, in Maxwell's electrodynamics). Particles are of non-zero spatial size and are `discrete' in that `being matter' is binary in this paradigm: a point in space and time either does or does not `contain' matter. Contrasting with a field, we could think of a $\Z_2$ valued `matter indicator function' on space and time (taking the value $0$ to represent the presence of matter), whereas fields are described in the `classical' paradigm by $\R$ valued functions. In both classical paradigms our `physical objects', whether particles or fields, exist and interact within (but not with) a `fixed' `box' or `background' of Euclidean space and time. We can think of particles, fields, space and time as the four `elements' of classical paradigms\footnote{Of the four `elements' space is perhaps fundamental in the classical paradigms since conceptually space could exist without particles, fields or time whereas the other elements require the existence of space.}, these elements co-exist and to an extent interact within classical theories, yet they are in an intuitive sense complementary rather than unified.

With wave-particle duality at its core, Quantum Mechanics can be seen as an attempt to unify the particle and field paradigms. Although it is experimentally highly successful, the `interpretation' of QM has proved obscure to the point that its study has become a field in its own right. The development of Quantum Field Theory has seen the identification of four fundamental forces, and the proposed unification of three of these forces into the Standard Model (SM). Subsequently much effort has been expended in the as yet unsuccessful attempt to formulate a quantum theory of gravity (Quantum Gravity, QG) and unify it with SM, thus incorporating the insights of GR into a QFT framework. QFT retains the interpretational difficulties of QM.

Crucially for what follows, QFT assigns discrete, `quantised' values to all of a particle's internal degrees of freedom, suggesting an intuition that a particle undergoes {\it`no internal changes'} or {\it`no internal evolution'} outside of `sudden' interactions.

Taking a pragmatic (and indeed almost instrumental) view of time, Special Relativity (SR) unifies space and time into a Minkowski `spacetime' in a manner that coherently integrates (or unifies) the viewpoints of all `internally related' observers. General Relativity builds on this to unify the perspectives of arbitrary observers in a manner that also unifies the gravitational field with spacetime, which becomes a Lorentzian curved manifold. Attempts to extend this framework to incorporate other forces have proved thus far unsuccessful, and GR remains a theory of gravity. Perhaps even more significantly GR remains an `incomplete' theory, in that even in its modelling of gravity GR requires an additional theory (which could be thought of as a `theory of matter') to elaborate on the stress-energy tensor, or to specify that particles should move on geodesics. The hoped for theory of Quantum Gravity might be looked to as a solution for this incompleteness, alternately the approaches referred to as `geometrodynamics' (which we will discuss below) look to resolve this issue within the GR framework before considering `quantization'.

\subsection{An Alternate Approach}\label{subsec:an alternate approach}

The search for a quantum theory of gravity takes QM as fundamental and seeks in QG to incorporate the insights of GR into the framework of QFT, and then to unify it with the Standard Model. This approach has yet to bear fruit, while the interpretational difficulties of Quantum Mechanics remain.

\begin{figure}[ht]
\centering
\includegraphics[scale=0.5]{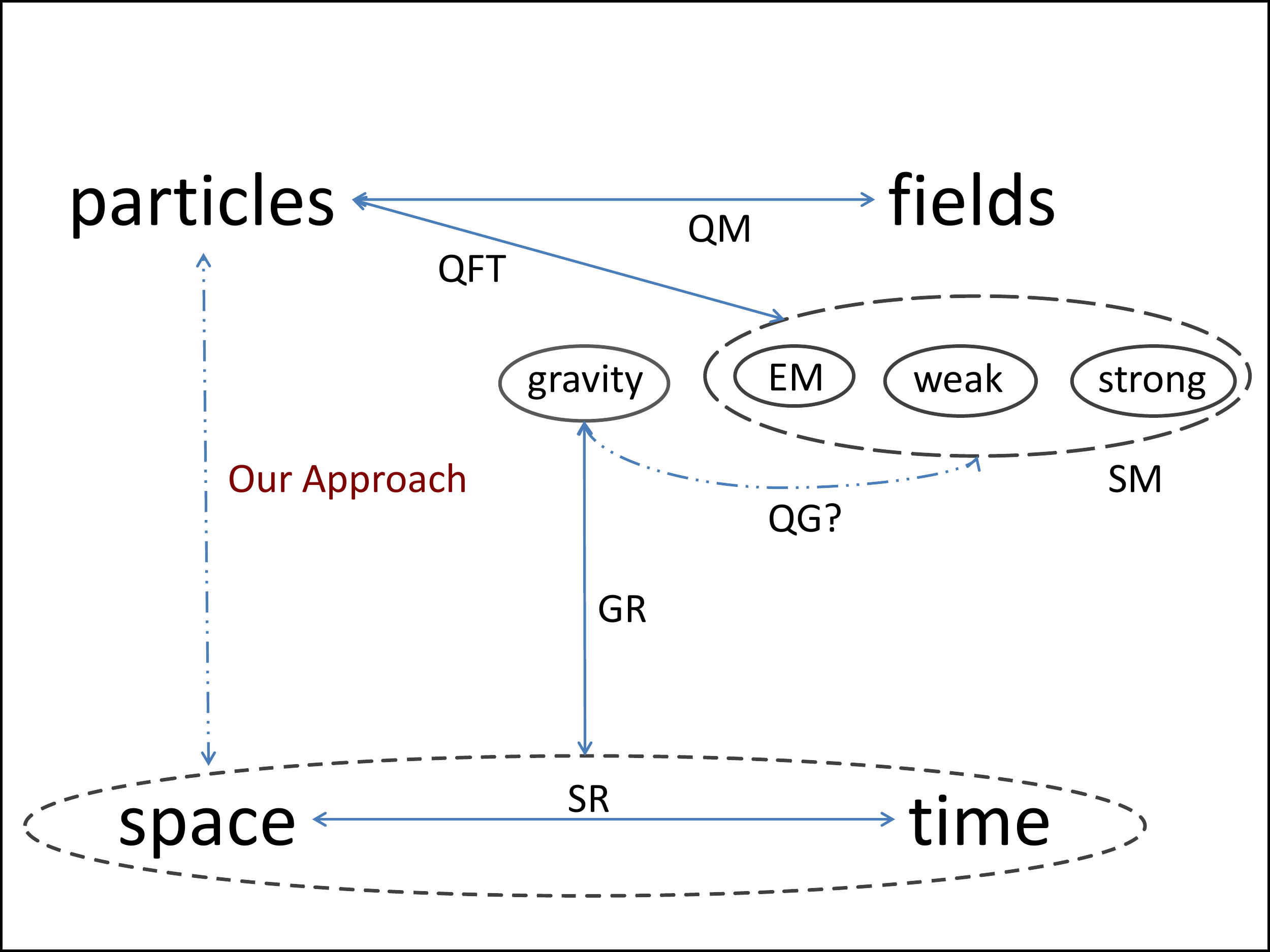}
\caption{An Alternate Approach}
\label{fig:diagram}
\end{figure}

We will adopt an alternate approach, seeking instead to unify the concept of discrete particles with that of GR's spacetime (figure \ref{fig:diagram}). As regards particles, we will take from the `classical' paradigm the notions of spatial discreteness and non-zero spatial size, and from QFT the intuition that a particle undergoes no internal evolution. As regards space and time, we will build on GR's spacetime framework, which neatly unifies space, time and the gravitational field. Our aim is to construct a framework which is `unified' in the sense that it unifies what we have called the four elements of classical physics, and `complete' in the sense that it does not require an external factor such as the stress-energy tensor in GR or a mechanism for `state-vector collapse' in QM. Because GR has already unified the gravitational field with spacetime our approach may most immediately lend itself to the formulation of a complete theory of gravity, however we hope that our framework will be flexible enough to also incorporate other forces, ideally within a single theory. We will leave to further research even conceptual thinking regarding the relationship between our approach and the QFT framework, including issues such as `quantization', in the interpretation of QM or the so called `paradoxes' of QM.

\subsection{Space without Time}\label{subsec:space without time}

There have been several attempts to develop General Relativity into a complete theory without (or before) `quantization' by seeking to describe all of physics with geometry, specifically the geometry of GR. This approach has been denoted as `geometrodynamics', most notably by Wheeler \cite{geometrodynamics:1961}, who we follow in describing the field. The idea itself can be traced back to Riemann \cite{RiemG:1867} (translated into English by Clifford \cite{RiemE:1873}), and is expressed clearly by Clifford \cite{Cliff:1876}, before being thrust into centre stage by Einstein's General Relativity. Indeed Einstein's matter free equation,
$$
R_{ab}=0,
$$
can be regarded as a complete theory of gravity. It is proposed as such in the `geon' model explored by Wheeler \cite{geometrodynamics:1961,geons:1957}, which identifies matter with spacetime curvature, for example in the form of `gravitational wave energy'. This differs from our idea of matter in lacking spatial discreteness and in allowing internal evolution inside matter. More generally, any specification of the stress-energy tensor would allow us to think of the Einstein equation as representing a complete theory, which we might seek to express in purely geometric terms\footnote{For example the dust solution $G_{ab}=\rho u^a u^b$ implies the purely geometric $R^a_b R^b_c -1/4 R^2 g^a_c=0$.}. However, as before spatial discreteness of matter would be lacking in such a theory. Einstein, Grommer, Infeld and Hoffmann \cite{EinGrom:1927,EinInfHoff:1938} perhaps come closer to our ideas, associating matter with singularities in the metric, which differs from our ideas in assigning zero spatial size to matter. We will develop the theory along a different path.

Our approach differs from the above in several ways. Firstly we will allow ourselves to alter General Relativity. Secondly, as discussed in the previous section, the formulation of a complete theory is only one aspect of our larger agenda. In particular we have quite specific ideas about the nature of matter, which we can crystalise into a requirement that particles are,
\bd
\item[P1] Spatially discrete and of non-zero spatial size.
\item[P2] Undergo no internal evolution.
\ed
Bearing in mind relativity's pragmatic approach to time, if there is to be no change inside a particle then we should conclude that no time passes either. This leads us to propose that matter is a region of space without time\footnote{Intuitively this fits in with our previous observation that space is the most fundamental of the four `elements' of classical physics.}.

\section{Mathematical Expression}\label{sec:mathematical expression}

\subsection{Basic Definitions}\label{subsec:definitions}

Since the idea of `space without time' suggests that matter is fundamentally three dimensional in nature, we might seek to model the universe by patching together four dimensional manifold(s) equipped with a Lorentzian metric (representing spacetime) and three dimensional manifold(s) with a Euclidean metric (representing matter). However, it seems simpler to retain a four dimensional manifold and incorporate matter by adjusting the metric.

We will partition the manifold into regions denoted as `spacetime', in which we will retain the basic GR structure\footnote{Further restrictions such as causality conditions can also be applied, and may be of interest to us, but we will not consider them here.}, and regions denoted as `matter', in which we will introduce new behavior\footnote{We may wish to restrict pathologies by constraining spacetime or matter regions to be of non-zero `size' or measure in some sense.}. As in GR, the geometry everywhere will be described by a $(0,2)$ tensor field $g$ which we call the metric and which we will in general assume to be smooth, though we will retain the formal flexibility to assume otherwise. In spacetime regions $g$ will be the usual GR metric, but in matter regions it will display new behavior. Since the manifold as a whole can no longer be referred to as spacetime we will denote it as `the universe' or more mundanely `the manifold'.

Looking back to the discussion of section \ref{sec:introduction}, we went from the idea of `no change inside matter' to that of `matter as space without time'. Though we took this step because our pragmatic view of time suggested these statements are physically equivalent, in terms of the technicalities of our mathematical framework we express them as two separate assumptions. Inside matter we want,
\ben
\item Space but no time.
\item No change.
\een

Starting with the first assumption, we note that at a spacetime point $p$ the Lorentzian metric separates the tangent space $\TM$ into {\it spacelike}, {\it timelike} and null {\it lightlike} vectors, so that a change of basis allows the metric to take the Minkowski form\footnote{Though not necessarily at more than one point simultaneously} in $\TM$
$$
\begin{pmatrix}
-1 & 0 & 0 & 0 \\
0 & 1 & 0 & 0 \\
0 & 0 & 1 & 0 \\
0 & 0 & 0 & 1
\end{pmatrix}.
$$
We will require that at any point $p$ inside matter there is a change of basis in $\TM$ which allows the metric to be written in the following canonical form,
$$
\begin{pmatrix}
0 & 0 & 0 & 0 \\
0 & 1 & 0 & 0 \\
0 & 0 & 1 & 0 \\
0 & 0 & 0 & 1
\end{pmatrix},
$$
so that instead of one negative and three positive eigenvalues the metric now has one null and three positive eigenvalues. We will denote null vectors inside matter (and the directions they define) as `{\it matterlike}'. Note that the metric is degenerate inside matter.

We now turn to the second assumption. Since the thrust of our ideas is to unify existing structures into space, and since we are proceeding from the GR framework in a region with no time, we interpret `no change' to mean `no change in the spatial metric', which in consideration of the above is equivalent to `no change in the metric'. We will adopt what is perhaps the most intuitive means of expressing this concept more precisely, requiring the Lie derivative of the metric in a matterlike direction to be zero,
$$
\L_v g=0,
$$
whenever $v$ is matterlike. We can formalise this discussion into the following definition.
\bD\label{def:matter}
Given a four manifold $\M$ with a $(0,2)$ tensor field $g$ which we denote the metric tensor, we say that a point $p\in\M$ is `\textbf{inside matter}' if the following two conditions are satisfied:
\bd
\item[Matter Assumption 1:] There exists a basis $\{e_0,e_1,e_2,e_3\}$ of $\TM$ such that \linebreak $g(e_a,e_b)=0$ for $a\neq b$, $g(e_0,e_0)=0$ and $g(e_i,e_i)=1$ for $i\in\{1,2,3\}$.
\item[Matter Assumption 2:] $g(v_p,v_p)=0\Rightarrow \L_v g |_p=0$.
\ed
Further we say that $p\in\M$ is `\textbf{in spacetime}' if the following holds:
\bd
\item[Spacetime Assumption:] There exists a basis $\{e_0,e_1,e_2,e_3\}$ of $\TM$ such that \linebreak $g(e_a,e_b)=0$ for $a\neq b$, $g(e_0,e_0)=-1$ and $g(e_i,e_i)=1$ for $i\in\{1,2,3\}$.
\ed
A region $R\subset\M$ is denoted as `\textbf{matter}' if every $p\in R$ is inside matter and `\textbf{spacetime}' if every $p\in R$ is inside spacetime. We say that the metric is \textbf{matterlike} inside matter and \textbf{Lorentzian} in spacetime. We call the manifold an `\textbf{MST manifold}', an `\textbf{MST universe}' or simply a `\textbf{universe}' if every point $p$ is either inside matter or in spacetime\footnote{We may want to restrict pathologies by requiring matter and spacelike regions to be of non-zero `size'; however we will not consider this here.}.
\eD
We can increase our understanding by considering a simple example.
\bE\label{example:simple particle}
Let $\M$ be a four manifold with the same topology and differential structure as the standard $\R^4$. We can place a global coordinate system on $\M$ in the standard fashion, with coordinates $\{t,x,y,z\}$, and equip $\M$ with a metric $g$ which in this coordinate system takes the form:
\beq
g=
\begin{pmatrix}
m(r) & 0 & 0 & 0 \\
0 & 1 & 0 & 0 \\
0 & 0 & 1 & 0 \\
0 & 0 & 0 & 1
\end{pmatrix},
\eeq
where $r=(x^2+y^2+z^2)^{1/2}$. Setting $m(r)=-1$ would yield the Minkowski metric, while setting $m(r)=0$ would yield matter. We can join these two behaviors,
$$
m(r)=\left\{\ba{cc} 0 & r\leq 1 \\ -e^{-(r-1)^{-2}} & r>1 \ea\right.
$$
Noting that $\partial_t$ is a Killing field (so that $\L_{\partial_t}g=0$ everywhere), it is easy to see that $g$ is matterlike for $r\leq 1$ and Lorentzian for $r>1$, further $g$ is smooth though not analytic. It is the function $m$ which identifies, or {\it indicates}, the presence of matter, constituting a smooth real valued version of the $\Z_2$ valued {\it `matter indicator function'} we encountered in section \ref{subsec:through the prism of unification}. As we shall see matter indicator functions of this nature will play a crucial role in what follows.

\begin{figure}[ht]
\centering
\includegraphics[scale=0.5]{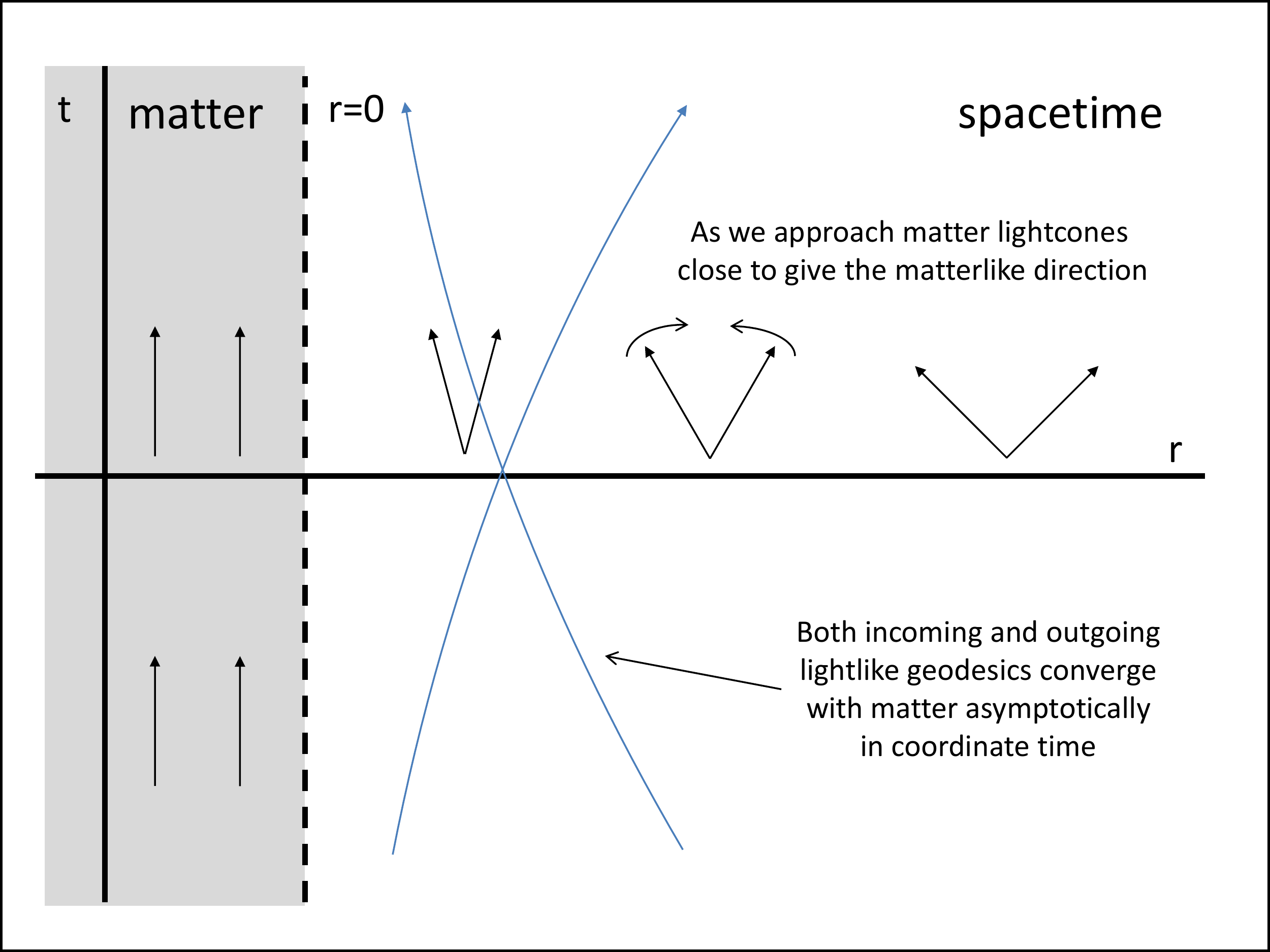}
\caption{Lightcones and Null Geodesics in Example \ref{example:simple particle}}
\label{fig:lightcones}
\end{figure}

We now examine the behavior of lightcones as we approach matter. Restricting to the $(t,x)$ $2$-plane, a null vector will have the form $v=\partial_t+a \partial_x$, and so,
\bea
0 &=& g(v,v)\nonumber \\
 &=& g(\partial_t,\partial_t)+a^2 g(\partial_x,\partial_x) \nonumber \\
 &=& m(r)+a^2 \nonumber\\
 \Rightarrow a &=& \pm(-m(r))^{1/2} \nonumber \\
 &=& \pm e^{-1/2(r-1)^{-2}}. \nonumber
\eea
Notice that $a\rightarrow 0$ and so $v\rightarrow \partial_t$ as we approach the $r=1$ boundary of matter from the right. Intuitively the three dimensional null lightcone `closes' to become the one dimensional matterlike direction, `squashing' the timelike direction in the process (figure \ref{fig:lightcones}). Notice that as we approach $r=1$ both incoming and outgoing lightlike geodesics `steepen' to asymptotically (in coordinate time) become matterlike and converge with the surface of the matter region.
\eE

The degeneracy of the metric inside matter presents us with some technical difficulties. The degeneracy of $g_{ab}$ leads to singular behavior in $g^{ab}$, which in turn leads to the singular behavior of connections and thus curvature terms inside matter. This is problematic since following General Relativity we would expect to formulate our theories in terms of curvature tensors such as the Ricci tensor. Even if we claim that the nature of matter regions should be regarded as a boundary condition and thus not in need of a governing equation, we will certainly wish to join spacetime regions to matter and so the behavior of curvature terms as we approach matter is of particular concern.

We explore these issues in the remainder of section \ref{sec:mathematical expression}. In section \ref{subsec:single observer} we tailor the standard ADM $3+1$ decomposition \cite{ADM:1962} to our needs, defining a general matter indicator function and use it to diagnose and classify the singular behavior near matter. In section \ref{subsec:the three observers} we perform three simultaneous $3+1$ decompositions corresponding to three observer's viewpoints, and see that we can fully describe the four geometry using spatial degrees of freedom. In section \ref{subsec:approaching matter} we compare the three observer's viewpoints and rebuild some of our four dimensional curvature terms so that they are well behaved near matter.

\subsection{A Single Observer: The 3+1 Framework}\label{subsec:single observer}

The degeneracy of the matrix inside matter is intuitively caused by the disappearance of time while space retains its `usual' nature, suggesting that we may gain some insight by examining a $3+1$ decomposition of the system in the manner of the ADM analysis \cite{ADM:1962}. Particular attention will be paid to the behavior of the geometry as we approach matter.

We start with a four manifold $\M$ equipped with a non-singular metric $g$ and a coordinate patch $U_X\subset\M$ which we will foliate with spacelike hypersurfaces $\S_{x^0}$ such that the spacelike coordinates $\{x^1,x^2,x^3\}$ parameterize the surface while $x^0$ is timelike and parameterizes the hypersurfaces. We denote the corresponding tangent space basis as $\{\d_a\}_{a=0}^3$, with $\{\d_i\}_{i=1}^3$ forming a tangent space basis for $\S_{x^0}$. We will denote by $\n$ the unit vector field normal to the hypersurfaces (and pointing in the positive $x^0$ direction), and the restriction of $g_{ab}$ to the hypersurface tangent bundle by $h_{ij}$, using $a,b,c,d$ to denote four dimensional indices and $i,j,k,l$ to denote three dimensional indices in the usual fashion.

The thrust of the $3+1$ decomposition is to express the geometry of the full four manifold in terms of the intrinsic and extrinsic geometry of the three dimensional hypersurfaces $\S_{x^0}$. In practise this comes down to $h$, $\n$ and derivatives thereof. However as we approach matter $g(\d_0,\d_0)\rightarrow 0$, which leads the components of the normal vector to become singular to preserve $g(\n,\n)=-1$. Thus for simplicity, and to further isolate the degeneracy of the metric to a single variable, we will write $\v=m\n$ where $m\in\R$ and $\v$ is chosen so that $\v^0=1$, implying $\n^0=m^{-1}$. The intention is for $\v^a$ to be finite everywhere so that $\v$ is timelike outside matter and matterlike inside matter. We think of $m$ as a `matter indicator function' as in example \ref{example:simple particle}, with $m=0$ `indicating' a matter region. As previously noted, the matter indicator function will prove to be a key concept in exploring the behavior of spacetime near matter. Having identified $m$, it is natural to use $m\rightarrow 0$ to describe the approach to a matter region, neatly avoiding the more involved detailing of this process in terms of spacetime paths. This further suggests that the behavior of geometric objects near matter be described in terms of power series in $m$. However, to be confident of this approach we must first check the coordinate dependence of $m$, which has been defined in terms of a particular coordinate system, to ensure that it `fully' captures the degenerate behavior of $g$. Specifically, we would want to check that an $m_Y$ defined using a different coordinate system $\{y^0,y^1,y^2,y^3\}$ should be of order one in $m$. We will leave this to the next section, but will nevertheless begin to explore the behavior of geometric structures (as expressed in the $\{x^0,x^1,x^2,x^3\}$ system) near matter in terms of $m$.

Naturally we start with $\v$ and the metric. From $g(v,\d_i)=0$ we can calculate the remaining components of $\v$ (where by $h^{ij}$ we mean the components of the inverse of the three dimensional matrix $h_{ij}$ and not $h_{ij}$ raised with $g$),
\bea
\v^0 &=& 1 \nn \\
\v^i &=& -\sum_{j=1}^3 g_{0j}h^{ji} \nn \\
\nn \\
\v_0 &=& -m^2 \nn \\
\v_i &=& 0. \label{eq:nu components}
\eea
Which confirms that the components of $\v$ are non-singular. Similarly we can decompose $g$ in terms of $v$, $m$ and $h$:
\bea
g_{00} &=& -m^2 +\sum_{i,j=1}^3\v^i \v^j h_{ij} \nn \\
g_{0i} &=& -\sum_{j=1}^3\v^j h_{ij} \nn \\
g_{ij} &=& h_{ij} \nn \\
\nn \\
g^{00} &=& -m^{-2} \nn \\
g^{0i} &=& -m^{-2} \v^i \nn\\
g^{ij} &=& h^{ij}-m^{-2}\v^i \v^j, \label{eq:3+1 g decomposition}
\eea
from which we can see that the singular behavior of $g^{ab}$ near matter is accounted for by the $m^{-2}$ term.

In the study of embedded surfaces the metric $h$ can be referred to as the {\it first fundamental form} describing the intrinsic geometry of the surface. No less important is the {\it second fundamental form}, $K_{ij}$, describing the surface's extrinsic geometry. The second fundamental form typically arises from the decomposition of the covariant derivative in $\M$ to components tangential and normal to the hypersurface, and thus can be used to construct four dimensional connections and curvature tensors in $\M$ from hypersurface connections and curvature tensors in a manner similar to our use of $\v$ in reconstructing $g$ from $h$ above. We can define the second fundamental form in a variety of ways, for example:
\bea\label{eq:definitions of K}
K_{ij} &=& -g(\nabla_i \n,\d_j) \nn \\ 
&=& g(\n,\nabla_i \d_j) \nn \\ 
&=& -m \G^{0}_{ij} \nn \\ 
&=& -\n_{(i;j)} \nn \\ 
&=& -1/2 (\L_n g)_{ij}.  \nn  
\eea
The simple relationship between $K$ and the derivative touches on the second matter assumption (definition \ref{def:matter}), but before we comment on this we will first note that a simple calculation of the connection term $\G^{0}_{ij}$ shows that:
\beq\label{eq:K in terms of v}
K_{ij} = -1/2 m^{-1}(\sum_{k=1}^3 h_{ik}(\d_j \v^k)+\sum_{k=1}^3 h_{jk}(\d_i \v^k)+\v(h_{ij})).
\eeq
Thus in general $K$ is $\ord{-1}$ and will become singular near matter. Now our second matter assumption can be written as $\L_{\v}g=0$ inside matter, which would imply that $\L_{\v}g\rightarrow 0$ as $m \rightarrow 0$. However we have as yet placed no constraint on how $\L_{\v}g$ approaches zero, as our matter assumptions address the behavior of the geometry inside, but not near, matter. It would seem natural to relate the limiting processes by which spacetime smoothly adopts our two matter assumptions as we approach a matter region, and so we introduce an additional assumption:
\bd
\item[The Derivative Assumption:] $\L_{\n}$ remains finite as we approach matter, $m\rightarrow 0$.
\ed
As this assumption was not a part of our original formulation we will highlight the places in which we adopt it, and note the effects of not doing so. Further, we will need to justify that this statement is coordinate independent, an issue we will address in the next section.

Finally we turn to curvature terms. Defining the {\it shape operator} $S_{ijkl}$ of the hypersurface in the usual way,
\beq\label{eq:shape operator definition}
S_{ijkl}=K_{ik}K_{jl}-K_{il}K_{jk},
\eeq
it can be shown that
\beq\label{eq:riemd 3+1 decomposotion}
R_{ijkl}=R_{Xijkl}+S_{ijkl},
\eeq
where $R_{Xijkl}$ is the intrinsic curvature of the hypersurface considered as an embedded manifold. Notice that this formulation only allows us to reconstruct the `spacelike' ($a,b,c,d\neq 0$) components of $R_{abcd}$. Now $R_{Xijkl}$ is defined solely in terms of $h_{ij}$ and is independent of $m$, and so will in general be $\ord{0}$. The order of $S_{ijkl}$ depends on the order of $K_{ij}$ in terms of which it is defined. Adopting the derivative assumption we have $S_{ijkl}$ and thus $R_{ijkl}$ being $\ord{0}$, whereas without the derivative assumption they are both $\ord{-2}$. We can extend this analysis to write,
\beq\label{eq:riemu 3+1 decomposotion}
R^{i}_{jkl}=R^{i}_{Xjkl}+S^{i}_{jkl}.
\eeq
Now $R^{i}_{Xjkl}= \sum_{p=1}^3 h^{ip}R_{Xpjkl}$ is the usual hypersurface curvature term, and is thus $\ord{0}$. However $S^{i}_{jkl}\neq \sum_{p=1}^3 h^{ip}S_{pjkl}$. Instead we have,
\bea
R^{i}_{jkl} &=& \sum_{a=0}^3 g^{ia} R_{ajkl} \nn \\
&=& g^{i0}R_{0jkl}+\sum_{p=1}^3 g^{ip}R_{pjkl} \nn \\
&=& -m^{-2}\v^i R_{0jkl} + \sum_{p=1}^3(h^{ip}-m^{-2}\v^i\v^p) R_{pjkl} \nn \\
&=& (R^{i}_{Xjkl} + \sum_{p=1}^3 h^{ip}S_{pjkl})-m^{-2}\v^i(R_{0jkl}+\sum_{p=1}^3\v^p R_{pjkl}), \label{eq:riemu series}
\eea
so that,
\beq\label{eq:Su definition}
S^{i}_{jkl}=\sum_{p=1}^3 h^{ip}S_{pjkl}-m^{-2}\v^i(R_{0jkl}+\sum_{p=1}^3\v^p R_{pjkl}),
\eeq
from which we can see that $S^{i}_{jkl}$, and thus $R^{i}_{jkl}$, are $\ord{-2}$ if we adopt the derivative assumption and $\ord{-4}$ if we do not. This suggests singular behavior in $R_{ab}$, which is of particular concern as we would like to build upon GR. A further problem is that our $3+1$ decomposition of curvature has not dealt with terms including non-hypersurface indices, for example $R_{0jkl}$. A new framework is needed.

\subsection{The Three Observers: The $3\times 3$ Framework}\label{subsec:the three observers}

Expecting the nature of space to be unaffected by the transition from spacetime to matter we have used the $3+1$ framework to express four dimensional tensors in terms of three dimensional spatial tensors. This has allowed us to isolate and classify singular behavior near matter (by order in $m$), but has not suggested a means of circumventing it. Further, the usual deconstruction of curvature terms applies only to components with purely spatial indices. A more powerful framework is needed.

We can think of the $3+1$ ADM framework as representing a single observer's viewpoint. In this formalism, an idealised observer $\Ob_X$ corresponds to a world line (segment) in spacetime, which forms the time `axis' in some physically derived coordinate system $\{x^i\}$ (for example `radar coordinates' \cite{radar1,radar2,radar3}) which cover a patch $U_X$ around the observer's world line (segment). As a single observer's viewpoint proves insufficient, we will deploy multiple observers. Bringing together the viewpoints of three observers allows us to ignore the timelike aspect of each system and provide a spatial `$3\times 3$' framework.

We begin with three observers, $\Ob_X, \Ob_y, \Ob_Z$ all passing though a common point $p_{XYZ}\in\M$, with corresponding coordinate\footnote{Much of this analysis, including key results, would still hold if we used non-coordinate basis frames. However we will not consider this generalisation here.} patches $U_X,U_Y,U_Z$ with a nonempty intersection $\tilde{U}_{XYZ}$. We undertake the $3+1$ ADM decomposition in each system, yielding hypersurfaces $\S_{X,x^0},\S_{Y,y^0},\S_{Z,z^0}$, whose normal vectors we will require to be independent at $p_{XYZ}$. We can then extend this to a connected neighbourhood $U_{XYZ}\subset\widetilde{U}_{XYZ}$ of $p_{XYZ}$ which we call a {\it tri-coordinate patch}\footnote{It may be desirable to place further conditions upon $U_{XYZ}$, or to take (in some systematic fashion) a subset thereof satisfying such constraints. We will leave this issue to later enquiry.}. We can think of an atlas of tri-coordinate patches covering our manifold. At each point $p\in U_{XYZ}$ the hypersurfaces define three dimensional spacelike linear subspaces of the tangent space, which we will denote $TX_p,TY_p,TZ_p$ (and refer to as {\it planes}), with intersections $TXY_p=TX_p\cap TY_p$ etc. We will use $A,B,C$ to denote an unspecified observer's system, so that $m_A$ is the matter indicator function corresponding to observer $\Ob_A$. Unless specifically mentioned otherwise we will henceforth assume that $p\in U_{XYZ}$.

\begin{figure}[htb]
\centering
\includegraphics[scale=0.5]{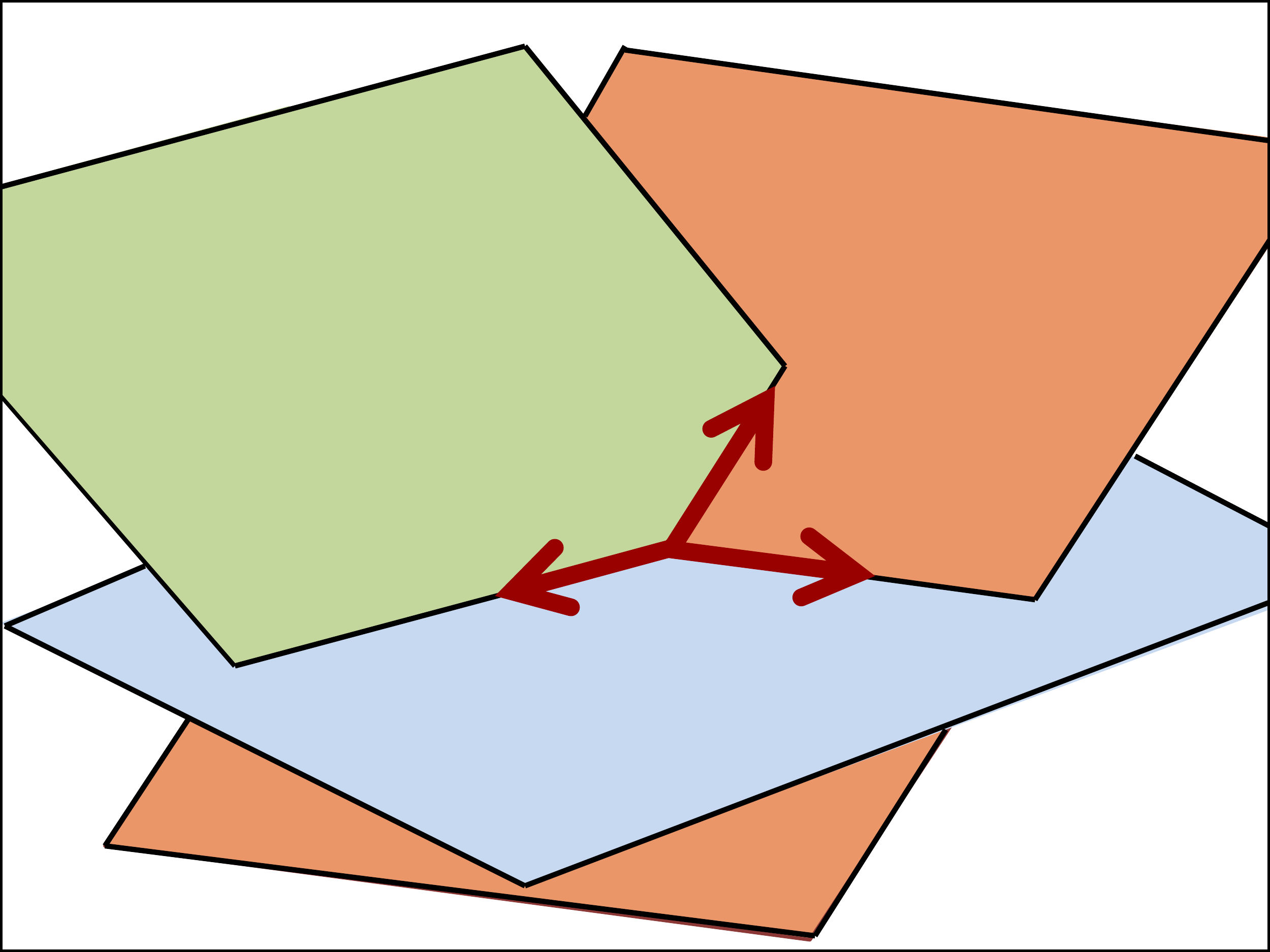}
\caption{The Intersection Basis in Three Dimensions}
\label{fig:intersection basis}
\end{figure}

Utilizing three observer's viewpoints simultaneously can be unwieldy, we therefore seek a means of bringing these viewpoints together. For conceptual simplicity we begin by considering a lower dimensional case, with three $2+1$ systems on a three dimensional manifold. Considering the tangent space $\TM$, we see that the three planes intersect at a point (the origin), and that each pair of planes $A,B$ intersects in a line ($TAB_p$ is one dimensional) which defines a unit vector $e_{AB}$, which is necessarily spacelike due to inclusion in $TA_p$ (in fact we have two unit vectors, and can choose the direction arbitrarily). Proceeding in this way we can define three spacelike unit vectors, whose independence is easy to establish from the independence of the $\n_A$, so that $\{e_{XY},e_{XZ},e_{YZ}\}$ is a basis, which we will call the { \it intersection basis} (figure \ref{fig:intersection basis}). Repeating this procedure at each point with suitable choice of directions for our basis vectors, we can extend the basis at one point to a basis frame.

Generalizing this procedure to four dimensions our three planes (now each a three dimensional subspace) in the tangent space $\TM$ intersect to give a line, with corresponding unit vector $e_{XYZ}$. Each pair of our planes now intersects in a two dimensional subspace $TAB_p$, which contains $e_{XYZ}$. From each $TAB_p$ we pick a unit vector $e_{AB}$ which is not co-linear with $e_{XYZ}$, yielding our spacelike intersection basis as before, which we extend to an intersection basis frame in some neighbourhood $U_{p}\subset U_{XYZ}$ around $p$. Note that the intersection basis does not necessarily integrate to yield a coordinate system. Notice also that the basis vectors $\{e_{XY},e_{XZ},e_{YZ}\}$ generate a fourth plane, $F$, which does not necessarily integrate to yield a hypersurface in $\M$.


For ease in expressing summations, it would be useful to adopt notation that allows us to enumerate the intersection basis vectors. We therefore write:
\bea
e_1 &=& e_{XYZ} \nn \\
e_2 &=& e_{XY} \nn \\
e_3 &=& e_{XZ} \nn \\
e_4 &=& e_{YZ}. \nn
\eea
We will continue to use indices $a,b,c,d$ to sum over $\{1,2,3,4\}$ and $i,j,k,l,p,q$ to sum over the three indices belonging to a particular plane. However, if we are considering $TY_p$ or $TZ_p$ the $i,j,k,l$ indices may for example take the value $4$. Thus $h_{Y14}=h_Y(e_{XYZ},e_{YZ})$. For clarity we will always write out the summation sign\footnote{We suggest a convention that when the summation sign is absent a sum over all four indices is assumed, however we will not use it here.}, and will denote the set of index values corresponding to $TA_p$ by $A$, so that,
\bea
X &=& \{1,2,3\} \nn \\
Y &=& \{1,2,4\} \nn \\
Z &=& \{1,3,4\}. \nn
\eea
Finally, we will write $A_{(i,j,k)}$ to mean a plane (there may be more than one possibility if there is degeneracy in $\{i,j,k\}$) containing the vectors $e_i,e_j,e_k$, and we will write $\overline{A}$ to denote the index value not in $A$, for example $\overline{X}=4$. As an example, note that the raising of an index for a three dimensional curvature term is written as follows,
\beq \nn
R^q_{A_{(i,j,k)} i,j,k} = \sum_{p\in A_{(i,j,k)}} h_{A_{(i,j,k)}}^{pq} R_{A_{(i,j,k)}p,i,j,k}.
\eeq

Using this notation, the intersection basis allows us to completely reconstruct the full four dimensional metric $g$ from the three dimensional spatial metrics $h_A$. Because of the overlapping nature of the three planes we have a choice of at least two $h_A$ with which to express each $g_{ab}$, for example $g_{12}=h_{X12}=h_{Y12}$. We here write out the metric in terms of the $h_A$, using the order of preference $X,Y,Z$ where there is a choice in the value of $A$,
\beq\label{eq:metric in intersection basis}
g=
\begin{pmatrix}
h_{X11} & h_{X12} & h_{X13} & h_{Y14} \\
h_{X21} & h_{X22} & h_{X23} & h_{Y24} \\
h_{X31} & h_{X32} & h_{X33} & h_{Z34} \\
h_{Y41} & h_{Y42} & h_{Z43} & h_{Y44}
\end{pmatrix}.
\eeq

Now given that geometric objects are all defined in terms of the metric, the fact that we can reconstruct the full four dimensional metric from the three dimensional `space' metrics suggests that we could simply shift entirely into a $3\times 3$ framework and write our equations using combinations of three dimensional terms, for example,
$$
R_{A ij} = 0,
$$
is a set of three equations that can be solved separately. The solutions of such equations would yield the three, three dimensional metrics which could be recombined into the full four dimensional metric\footnote{There would also be consistency conditions which would appear as constraint equations with the effect of enforcing $h_{Aij}=h_{Bij}$ where there are overlaps.}. However, we will stay with the usual four dimensional framework, and attempt to `recover' the use of the curvature terms that are of interest. The key to this will be in comparing the behaviors of the various planes as we approach matter, to which we turn in the next section.

\subsection{Approaching Matter}\label{subsec:approaching matter}

Starting with the viewpoint of $\Ob_X$, but using the intersection basis, we can perform a $3+1$ decomposition of the metric in terms of the hypersurfaces $\S_{X,x^0}$ as in section \ref{subsec:single observer}, but now using the spacelike $e_4$ rather than the timelike $e_0$ as the `non-tangent' basis vector. Adding an $X$ subscript to variables in the obvious fashion, it is easy to check that (\ref{eq:3+1 g decomposition}) still holds, and we can define $\v_X$, $K_X$ and so on as before. Extending this approach to all three observers, we can write down a plethora of objects, $\v_A$, $K_A$ and so forth. We will examine the relative behaviors of the $X$ and $Y$ objects as we approach matter, but since no observer is privileged this analysis will apply equally to any pair of observers.

Naturally we begin with the metric tensors. $TX_p$ and $TY_p$ share two of three basis vectors, leading their metrics to have substantial overlap.
\bea
h_Y &=&
\begin{pmatrix}
g_{11} & g_{12} & g_{14} \\
g_{21} & g_{22} & g_{24} \\
g_{41} & g_{42} & g_{44}
\end{pmatrix}
\nn \\
&=&
\begin{pmatrix}
h_{X11} & h_{X12} & -\sum_{j\in X}\v_X^j h_{X1j} \\
h_{X21} & h_{X22} & -\sum_{j\in X}\v_X^j h_{X2j} \\
-\sum_{j\in X}\v_X^j h_{X1j} & -\sum_{j\in X}\v_X^j h_{X2j} & -m_X^2+\sum_{i,j\in X}\v_X^i\v_X^j h_{Xij}
\end{pmatrix}
\label{eq:hy decomposition}
\eea
As we approach matter  $m_X\rightarrow 0$, so we define,
$$
h_{Ymat} = h_Y |_{m_X=0},
$$
so that,
\beq\label{eq:hY series hymat}
h_Y = h_{Ymat} - m_X^2 I_{44},
\eeq
where $(I_{44})_{ij}=\delta_{i4}\delta_{j4}$. From (\ref{eq:hy decomposition}) we can see that the degeneracy of $g$ inside matter leads $h_{Ymat}$ to resemble a basis transformation of $h_X$, explicitly inside matter $h_Y$ behaves as though,
\beq \nn 
e_4 = -\sum_{i\in X}\v_X^i e_i.
\eeq
Thus in terms of the geometry it appears as though the $TX_p$ and $TY_p$ planes converge, with $e_4$ `falling' into the $TX_p$ plane. Although in coordinate terms $e_4$ remains clearly distinct from $TX_p$, in terms of the metric it is as though $e_4$ literally falls into, and becomes part of, $TX_p$ as we approach matter, with the (metric) angle between $e_4$ and $TX_p$ decreasing to zero as $m_X\rightarrow 0$. This reflects the nature of our construction, faced with uniting fundamentally four dimensional spacetime and fundamentally three dimensional space in a single framework we chose to retain GR's four dimensional manifold (which gives us four dimensional topological and differential structure, including coordinate systems) while allowing the three dimensional nature of matter to be expressed by the metric. We will in what follows abuse notation and inside matter express $h_{Y}$ as if it were $h_X$ in a different basis, we hope it is clear that {\it such expressions apply to the components of $h_{Y}$  in the intersection basis}, and do not refer to $h_{Y}$ itself as a bilinear map from $TY_p$ to $\R$. The same applies for other geometric objects which we will treat in the same fashion.

We can make this `change in basis' more explicit, and write $h_{Ymat}=P^T h_X P$ where,
\beq \nn
P=
\begin{pmatrix}
1 & 0 & -\v_X^1 \\
0 & 1 & -\v_X^2 \\
0 & 0 & -\v_X^3
\end{pmatrix}
\eeq
which yields,
\beq\label{eq:hY series hx}
h_Y = P^T h_X P - m_X^2 I_{44}.
\eeq
This succinctly expresses the behavior of $h_Y$ near matter, as compared with $h_X$. We can go on to calculate similar comparisons for other terms. Writing $XY$ to represent $X\cap Y=\{1,2\}$, and $h_{XY}$ to represent the metric restricted to $TXY_p$, we have,
\beq\label{eq:dethy series}
det(h_Y) = det(h_{Ymat})-m_X^2 det(h_{XY}),
\eeq
and,
\beq\label{eq:invhy series}
h_Y^{-1} = (h_{Ymat})^{-1}+m_X^2(h_{Ymat})^{-1}I_{44}(h_{Ymat})^{-1} + \ordx{4}.
\eeq
Then using $\v_Y^i=-\sum_{j\in Y}g_{\overline{Y}j}h_Y^{ij}$ we find that,
\bea
\begin{pmatrix}
\v_Y^1 \\
\v_Y^2 \\
\v_Y^4
\end{pmatrix}
&=&
\frac{1}{\v_X^3}\begin{pmatrix}
\v_X^1 \\
\v_X^2 \\
1 \end{pmatrix}
+\ordx{2},          \nn
\eea
which can be written in our succinct notation as,
\beq \nn
\v_Y^i = \frac{1}{\v_X^{\overline{Y}}}\v_X^i +\ordx{2} \ \ \text{for} \ \ i\in Y,
\eeq
which can be generalized to,
\beq \nn
\v_B^i = \frac{1}{\v_A^{\overline{B}}}\v_A^i +O(m_A^2) \ \ \text{for} \ \ i\in B,
\eeq
and since $1=\v_B^{\overline{B}}=\v_{A}^{\overline{B}}/v_{A}^{\overline{B}}$ we can recover all of $\v_B$ in this way,
\beq \label{eq:vA series}
\v_B^a = \frac{1}{\v_A^{\overline{B}}}\v_A^a +O(m_A^2),
\eeq
which will be crucial later.

Turning to the matter indicator functions, we see that,
\bea
m_Y^2 &=& -g(\v_Y,\v_Y) \nn \\
&=& \frac{m_X^2}{(\v_X^3)^2} \frac{det(h_{Ymat})}{det(h_Y)} \nn \\
&=& \frac{m_X^2}{(\v_X^3)^2} + \ordx{4}, \nn
\eea
which generalizes to,
\beq\label{eq:ma series}
m_B^2 = \frac{m_A^2}{(\v_A^{\overline{B}})^2} + O(m_A^4).
\eeq
Thus $m_A$, $m_B$ are of the same order as we approach matter. Since the coordinate systems $\{x^a\}$, $\{y^a\}$, $\{z^a\}$ were arbitrary and pairwise unrelated, we have shown that the matter indicator function we have here defined is coordinate independent `up to order', and so is a meaningful measure of the `order' of structures as we approach matter. We have shown,
\bL\label{lemma:m coord independence}
The matter indicator function $m_X$ defined as above is coordinate independent (and so independent of the observer $\Ob_X$) up to order.
\eL

By using similar calculations, or by considering,
$$
K_{Y4i} = g(\n_Y,\nabla_{e_4}e_i),
$$
in light of the above, it is easy to show that,
$$
K_{Y4i} = \sum_{j\in X}\v_X^j K_{Xij} + \ordx{0} \ \ \text{for} \ \ i\in XY,
$$
and thus that
\beq\label{eq:K series}
K_Y = P^T K_X P + \ordx{0},
\eeq
so that the normal structures converge to the lowest order in the same fashion as the metric structures. This also means that $K_Y$ and $K_X$ are of the same order in $m_X$ (or indeed in $m_A$ for any $A$), thus from (\ref{eq:definitions of K}) we can conclude that the derivative assumption is well defined,
\bL\label{lemma:derivative assumption is well defined}
Using the notation defined above, the order of $\L_{n_{X}}g$ in $m_X$ is coordinate independent (and so independent of the observer $\Ob_X$), so that the derivative assumption is well defined and its validity independent of the choice of coordinates.
\eL

Continuing in the fashion we can show that,
\beq\label{eq:SY series}
S_{Yijkl} = P^{\overline{i}}_i P^{\overline{j}}_j P^{\overline{k}}_k P^{\overline{l}}_l S_{X\overline{i}\,\overline{j}\,\overline{k}\,\overline{l}}+\ordx{0},
\eeq
for $i,j,k,l \in Y$ and $\overline{i},\overline{j},\overline{k},\overline{l}\in X$.

Now returning to the decomposition of the curvature (\ref{eq:riemd 3+1 decomposotion}), we recall that in the absence of the derivative assumption the divergence of $R_{ijkl}$ near matter arises from the $\ordx{-2}$ behavior of the shape operator, whereas the intrinsic curvature is $\ordx{0}$. Using (\ref{eq:SY series}) we can generate combinations of Riemann curvature components in which the shape operator terms cancel, and are thus $\ordx{0}$. We would however still have to deal with the additional $m_X^{-2}$ factor introduced when we raise an index, which must be done to construct the Ricci tensor. We will for now adopt the derivative assumption, and turn our attention to raising an index and constructing the Ricci curvature.

Writing our $3+1$ breakdown of $R^i_{jkl}$ (\ref{eq:riemu series}) in our $3\times 3$ notation, for $i,j,k,l\in X$ we have,
\beq \nn
R^i_{jkl}= \sum_{p\in X}h_X^{ip}R_{Xpjkl} + (\sum_{p\in X} h_X^{ip}S_{Xpjkl}-m_X^{-2}\v_X^i(R_{4jkl}+\sum_{p\in X}\v_X^p R_{pjkl}))
\eeq
Focusing on the $\ordx{-2}$ term (the divergent `problem' term), we notice that $(R_{4jkl}-\sum_{p\in X}\v_X^p R_{pjkl})$ has no $i$ dependence, and is not summed with the $\v_X^i$. This suggests a way forward; for $r,s,i,j,k,l\in X$,
\beq \nn
\v_X^rR^s_{jkl} = \v_X^r \sum_{p\in X} h_X^{sp}R_{pjkl}-m_X^{-2}\v_X^r \v_X^s (R_{4jkl}+\sum_{p\in X}\v_X^p R_{pjkl}),
\eeq
has an $\ordx{-2}$ term which is symmetric in $r,s$, so that,
\beq \nn
\v_X^r R^s_{jkl} + \v_X^s R^r_{jkl} =\sum_{p\in X}(\v_X^r h_X^{sp} - \v_X^s h_X^{rp}) R_{pjkl}.
\eeq
Using the derivative assumption, we can generalize this to,
\beq \nn
\v_{A_{(j,k,l)}}^r R^s_{jkl} -\v_{A_{(j,k,l)}}^s R^r_{jkl} = O(m_{A_{(j,k,l)}}^0),
\eeq
where we must assume that $r,s\in A_{(j,k,l)}$. However if we choose $r=l$, $s=k$ then this condition is automatically satisfied. Further, as we have shown (\ref{eq:vA series}) that the $\v_A$ are of the same order in $m_X$, and that the $m_A$ are of the same order (\ref{eq:ma series}), we can state that (with no summation),
\bea
\v_B^l R^k_{jkl} - \v_B^k R^l_{jkl} &=& \ordc{0} \ \ \ \text{(no summation)}\nn \\
\Rightarrow \v_B^l R^k_{jkl} + \v_B^k R^l_{jlk} &=& \ordc{0} \ \ \ \text{(no summation)}, \nn
\eea
for any $B,C$. Since the choice of $B$ is no longer tied to $j,k,l$ we are free to sum the $k$ and $l$ indices over the whole space (ie from $1$ to $4$), yielding,
\bea
\sum_{1\leq l\leq 4}\v_B^l R_{jl} + \sum_{1\leq k\leq 4}\v_B^k R_{jk} &=& \ordc{0} \nn \\
\Rightarrow \sum_{1\leq k\leq 4}\v_B^k R_{jk} &=& \ordc{0}. \nn
\eea
We have just proved,
\bT\label{thm:vRic}
Let $\M$ be a universe obeying the derivative assumption, and let $\Ob_X$, $\Ob_Y$, $\Ob_Z$ be three observers whose worldlines intersect at $p_{XYZ}$. Then at any point $p\in U_{XYZ}$ we have,
$$
\sum_{1\leq a\leq 4}\v_A^a R_{ab} = \ordc{0}
$$
\eT

Now defining $H_A^{ab}$ to be the four matrix whose components are $h_A^{ab}$ for $a,b\in A$ and zero otherwise, we can write (\ref{eq:3+1 g decomposition}) as,
\beq \nn
g^{ab} = H_A^{ab} - m_A^{-2}\v_A^a \v_A^b,
\eeq
which yields,
\bea
R &=& \sum_{1\leq a,b\leq 4}H_A^{ab}R_{ab}-m_A^{-2}\sum_{1\leq a,b\leq 4}\v_A^a \v_A^b R_{ab} \nn \\
&=& O(m_A^{-2}), \nn
\eea
using Theorem \ref{thm:vRic}. Now noting that $\sum_{1\leq b\leq 4}g_{ab} \v_A^b$ is either zero or $-m_A^2$ and thus $\orda{2}$, we see that,
\bea
\sum_{1\leq a\leq 4}\v_A^a G_{ab} &=& \sum_{1\leq a\leq 4}\v_A^a R_{ab} + (\sum_{1\leq a\leq 4}\v_A^a g_{ab}) R \nn \\
&=& \orda{0}.
\eea
We have proved,
\bC
Let $\M$ be a universe obeying the derivative assumption, and let $\Ob_X$, $\Ob_Y$, $\Ob_Z$ be three observers whose worldlines intersect at $p_{XYZ}$. Then at any point $p\in U_{XYZ}$ we have,
$$
\sum_{1\leq a\leq 4}\v_A^a G_{ab} = \ordc{0}.
$$
\eC

These results are quite a surprise, the curvature tensors defined on the various planes recombining beautifully to reconstruct Ricci and Einstein curvature while neatly cancelling out the problem terms. This means that we can use a minor tweak on the familiar four dimensional Ricci and Einstein tensors to formulate equations that hold as we approach (and enter) matter, and so to construct `complete' theories in our unified matter-space-time framework.

\section{Summary and Discussion}\label{sec:conclusion}

\subsection{Summary}\label{subsec:summary}

While Quantum Gravity remains elusive and Quantum Field Theory retains the interpretational difficulties of Quantum Mechanics, we have introduced an alternate approach to the unification of particles, fields, space and time, suggesting that the concept of matter as space without time provides a framework which unifies matter with spacetime and in which we anticipate the development of complete theories (ideally a single unified theory) describing observed `particles, charges, fields and forces' solely with the geometry of our matter-space-time universe.

Formalizing our idea of matter in definition \ref{def:matter}, we introduced the matter indicator function $m$ which not only `indicates' the presence of matter but further allows us to describe the behavior of geometric objects as we approach matter, classifying them by their order in $m$. We encountered some technical difficulties, with the degeneracy of the metric leading some important geometric objects to become singular near matter. Noting that the geometry of spacelike hypersurfaces remains well behaved near matter we saw that the simultaneous use of three observers' viewpoints provides us with sufficient spatial degrees of freedom with which to describe the full geometry, with dynamical equations expressed in terms of the geometries of the three spacelike hypersurfaces corresponding to our observers. We then introduced the intersection basis, which proves to be a powerful tool, allowing us firstly to show the coordinate independence of the matter indicator function up to order and secondly to find the surprising and beautiful recombination of curvatures from the three hypersurfaces to provide versions of the usual Ricci and Einstein tensors that are well behaved near matter, allowing us to press forward to construct complete theories in this new framework using familiar geometric objects.

\subsection{Looking Ahead}\label{subsec:looking ahead}

Looking ahead to anticipated future theories, we briefly discuss some ways in which particles, charges, fields and forces might operate within our framework. We then comment on cosmology before concluding.
\bd
\item[\emph{Particles:}] Starting with an observer $\Ob_X$ and a matter region $R$ we call the intersection $R\cap \S_{X,x^0}$ {\it a particle in $\S_{X,x^0}$} (or {\it a particle in $X$ at $x^0$}) if it is connected and `surrounded' by a region of spacetime in $S_{X,x^0}$ (and thus in $U_X$). We will say that $R$ is {\it a particle\footnote{We may want to loosen the definition of a particle to include `fixed' three geometries moving in a lightlike direction, or in other words `beams' of lightlike Killing fields.} according to $\Ob_X$}, or alternatively {\it a particle in the $X$ system} or more succinctly {\it a particle in $X$}, if it is a particle in every $\S_{X,x^0}$. In what follows we will use the term `particle' loosely, implicitly assuming the observer $\Ob_X$.
\item[\emph{Charges:}] Our framework looks to geometry to explain the universe, thus a particle's `internal degrees of freedom' must be characterized by its three geometry, which is unchanging by assumption. Thus whatever `charge' (for example `rest mass') the particle's interior geometry encodes will thus be constant, and will not dissipate or collapse inwards. If a particle's interior geometry is homogeneous, it and thus the charges it encodes might be characterized by the Lie group of the particle's interior three geometry, which is reminiscent of QFT.
\item[\emph{Fields:}] The `fixed' geometry inside (more accurately at the boundary of)\footnote{In fact, the `interior' of a particle may have no contact with spacetime at all.} a particle will act as a boundary condition for spacetime, leading to approximately predictable patterns of spacetime geometry around the particle. The particle's effect will in general only be approximate as there may also be other influences on spacetime such as other particles or initial conditions. We denote a particle's influence on the surrounding spacetime as a \emph{field}.
\item[\emph{Forces:}] Consider two particles as seen by a non-accelerating observer $\Ob_X$, and assume that the minimal spacelike geodesics joining the two particles in each $\S_{X,x^0}$ lie entirely within $U_X$, and entirely within spacetime. The relative acceleration of the two particles is then a feature of the spacetime geometry in $U_X$, and we can see that the internal geometries of the two particles will affect the spacetime geometry of $U_X$ and thus the relative acceleration of the two particles. In this way, we see how charges lead to fields which lead to forces in our framework.
\item[\emph{Cosmology:}] We simply note that the introduction of matterlike regions allows for previously unconsidered cosmological structures. One striking example is a universe that is matterlike everywhere other than in a connected spacetime region of finite volume which is bounded on all sides (including the timelike direction) by matter. Particles in this universe might be thought of as `world tubes' connecting the `start of time' and `end of time' boundaries.
\ed

\chapter[Toward a Complete Theory of Gravity]{Toward a Complete Theory of Gravity}\label{ch:CG}

\begin{small}
\begin{center}
   \textbf{Abstract}
  \end{center}
   In this chapter we take the first steps toward a complete theory of gravity in the Matter-Space-Time framework, developing a governing equation by consideration of a static, spherically symmetric `Schwarzschild' scenario  with a central stationary spherical particle. We reject the matter free Einstein Equation on the grounds that it can not smoothly join matter and spacetime regions, instead building a simple toy model based on representing the `mass field' with the total curvature of the particle's three geometry. We use the insight gained from this toy model to propose a new governing equation for a complete theory of gravity.
\end{small}
\vskip 0.5cm

\section{First Steps}\label{sec:first steps}

In this chapter we take the first steps toward a complete theory of gravity in the Matter-Space-Time framework. Such a theory should smoothly join matter and spacetime regions, describing all gravitational effects without recourse to any mechanism external to the theory. We hope that this theory of \emph{Complete Gravity} (CG) will eventually reproduce the experimentally verified predictions of General Relativity, and expect it to approximately reproduce GR at the scales at which GR has been successfully tested. We will of course be curious to explore the differences between CG and GR  at larger and smaller scales, not least because such differences might allow for falsifiable prediction and experimental verification. Ultimately, we hope that  CG will become a feature of a broader unified theory with a single `\emph{underlying equation}' which describes all observed particles, charges, fields and forces.

For now we make a start by analysing theories in what is perhaps the simplest useful scenario, a universe with the topological and differentiable structure of the standard $\R^4$ containing a single, spherical, non-accelerating, non-rotating particle of matter, where the (smooth and smoothly joined) geometries of the matter and spacetime regions are as simple, plain and featureless as the theory allows. We begin by assuming that the geometry of the particle is homogeneous, so that it can be characterized by a Lie group as alluded to in section \ref{subsec:looking ahead}. Since, as far as we are aware, gravity is unipolar and does not imply a preferred direction, it seems natural to impose the symmetry group $O(3)$ so that the geometry of the particle is isotropic, which in this case means it is spherically symmetric. Turning to our spacetime, it is now natural to assume spherical symmetry, and for additional simplicity we will further assume that spacetime is static. Thus our picture is of a single, spherical, non-accelerating, non-rotating particle with spherically symmetric internal geometry joining smoothly with a static, spherically symmetric spacetime, which we can perhaps think of as being `at equilibrium'.

Now to concretely make predictions we would have to model an experimentally realisable (or at least physically observable) scenario, for example by modelling two physical particles in a manner that is experimentally realisable, or by approximating the behaviour of large quantities of particles and looking to cosmology. However we make a start by requiring the behaviour of the spacetime in our simple model to asymptotically match the behaviour described by GR. Due to the static spherical symmetry we have assumed this means we require our spacetime to be asymptotically Schwarzschild.

Note that even in regions where spacetime behaviour matches that predicted by General Relativity we are not guaranteed to see physically observable behaviour matching the predictions of GR. For example, GR assumes that particles move along geodesics, so that a relatively small particle near a much larger mass will approximately follow a Schwarzschild geodesic. However in our framework the picture may be more complicated, since the small particle will be matter rather than spacetime with as yet unexplored consequences for the two-body problem. Further while the geometry of the spacetime does give us the structure of lightlike geodesics this may well differ from the behaviour of observable light, which in our framework may for example consist of particles of matter or `lightlike particles' (beams of lightlike Killing fields) in which the fixed three geometry encodes `colour' and `brightness', and which may only approximately travel along lightlike geodesics, or whose presence may alter the geodesic structure. To make concrete predictions concerning light we may first need to construct a `complete theory of electrodynamics' in our Matter-Space-Time framework.

With these caveats made we may proceed. In section \ref{sec:toward a complete theory of gravity} we develop a governing equation for a complete theory of gravity in the MST framework. We begin in section \ref{subsec:Ric=0} by considering the matter free Einstein Equation, which we reject due to its inability to smoothly join matter and spacetime regions. In section \ref{subsec:toy model} we begin the construction of a toy model universe with a static, spherically symmetric metric and a stationary, non-rotating spherical particle centered at the `origin', whose three geometry we model as a space of constant positive curvature. We describe the spacetime of our toy model using a single matter indicator function which smoothly joins matter to spacetime and yields asymptotic Schwarzschild behaviour. In section \ref{subsec:curvature in toy model} we propose an explicit matter indicator function yielding a concrete metric, we then examine the spatial `density' related curvatures followed by the `acceleration' related curvatures which involve a time direction. In section \ref{subsec:equation} we build on the insights gained from our intuitive toy model to propose our new governing equation. Section \ref{sec:conclusion} summarises and looks ahead to areas of potential future research.

\section{Toward a Complete Theory of Gravity}\label{sec:toward a complete theory of gravity}

\subsection{Einstein's Empty Space Equation}\label{subsec:Ric=0}

The Matter-Space-Time framework takes its conception of spacetime from General Relativity, in which the gravitational field comes neatly `pre-unified' with the spacetime `background' making it natural for us to begin by examining Einstein's Equation. As previously mentioned, to make General relativity a complete theory we must specify the stress-energy tensor $T_{ab}$. Since $T_{ab}|_p$ represents the presence of matter at $p$ it seems natural to try the `empty space' Einstein Equation,
\beq\label{eq:Ric=0}
R_{ab}=0,
\eeq
in spacetime, and now describe matter using our `space without time' matter regions rather than using a non-zero stress-energy tensor in spacetime. For (\ref{eq:Ric=0}) to make sense as we approach matter we recast it in an intersection basis,
\beq\label{eq:RicAb=0}
R_{Ab}=0.
\eeq
This form of the equation can hold everywhere (in both matter and spacetime regions). Recall that we can recover $R_{ab}$ from $R_{Ab}$ in spacetime, so (\ref{eq:Ric=0}) holds wherever it is well-defined.

We now look into the `one-body problem', enquiring if (\ref{eq:RicAb=0}) admits solutions conforming to the static, spherically symmetric scenario outlined above. We start by assuming that in this scenario we can find a smooth metric, $g_{sol}$, which yields matter for $r\leq r_0$ and spacetime for $r>r_0$. Then in the spacetime region $r>r_0$ the metric is non-degenerate and so by Birkhoff's theorem our assumption of spherical symmetry forces $g_{sol}$ to coincide with a Schwarzschild metric, $g_{Sch}$, parameterized by a constant $M$ which is physically interpreted as the mass of static, spherical matter centered at $r=0$. Now as the metric is degenerate inside matter and smooth everywhere it must smoothly become degenerate as we approach $r=r_0$ from the right. However since the Schwarzschild metric can be extended in the region $r>0$ (using, for example, Lemaître coordinates) and is nowhere degenerate, its determinant can not tend to zero as we approach $r_0$ from the right meaning that we must have a discontinuity at the boundary of matter contradicting our requirement that $g_{sol}$ be smooth. We must therefore reject the matter free Einstein Equation\footnote{Since the general Einstein Equation with an unspecified stress-energy tensor carries essentially no information, any equation we adopt could be framed as an Einstein Equation with a suitable choice of $T_{ab}$. However, once we reject the matter free equation we no longer find it useful to think within this framework.} and look to a new theory.

\subsection{A Toy Model}\label{subsec:toy model}

To guide us in constructing a new theory we will first consider a toy model metric, $\g$, in the simple static, spherically symmetric scenario outlined above. We will base our model on our physical intuition, and hope that this `fleshed out' example will help us gain insight and intuition regarding the behaviour of gravity in this scenario, so that we are then able to suggest a governing equation leading to a new theory of gravity.

To re-iterate our scenario, we take a manifold $\M$ with the same topological and differential structure as the standard $\R^4$, and use the usual spherical coordinate chart $\{t,r,\th,\phi\}$. We equip $\M$ with a static, spherically symmetric metric $\g$, which gives us a matter region $r\leq r_0$ for some $r_0>0$, and spacetime when $r>r_0$. Stationarity automatically implies that $\d_t$ is a Killing field everywhere, so matter assumption $2$ is automatically satisfied, as is the derivative assumption, and we need only be concerned with matter assumption $1$; which we can achieve by requiring $\gtt=0$ for $r\leq r_0$.

We begin by considering our particle, the matter region $r\leq r_0$. As before we assume homogeneity and isotropy to yield a space of constant curvature, though we note that a physical particle this simple can not necessarily be found in nature. Now let us apply some physical intuition; classically the `charge' associated with gravity is mass, and we could think of a `mass field' assigning a positive real number, the `mass density' to each point in spacetime. Intuitively, in this picture mass can be thought of as a simple measure of the quantity or `amount' of matter. If one region in a spacelike hypersurface contains `more mass' than a second region, we think of the first region as in some sense containing more matter. Thus classically, in a given hypersurface we can think of mass density as the `density of matter'. Switching now to our matter-space-time framework we can no longer make recourse to an external field and must describe our intuition of mass density using the three geometry of a particle. Then the most direct and natural translation of `amount of matter' is the `amount of space', with the idea of mass (or matter) density being expressed as volume density. To make this more precise, consider an observer $\Ob_X$ with associated spacelike surfaces $\Sx$. We can measure the `size' of a spherical region in $\Sx$ by its surface area, and the `amount' of space inside by the volume of the spherical region. Thus comparing the volume to the surface area gives us an intuitive notion of density, and we thus expect the volume to surface area ratio to be larger inside a particle with mass than it is in the Minkowski metric of `pure spacetime'. Letting the radius of our sphere shrink to zero so that we can recover a notion of mass density at a point, our volume to surface ratio leads us to the total sectional curvature of the hypersurface,
\beq\nn
K_S = K_{12}+K_{13}+K_{23},
\eeq
where,
\beq\nn
K_{ij} = \frac{R_{ijij}}{g_{ii}g_{jj}-g_{ij}^2},
\eeq
is the sectional curvature of the $\{x^i,x^j\}$ $2$-plane. We call $K_S$ the \emph{density of $\Sx$}, inside matter we can simply refer to it as the \emph{density}. Now noticing that in a space of constant curvature $K_S$ is a constant $K_{mat}$ ($K_S$ \emph{is} the constant curvature), the results of our physical intuition match our Lie algebra based reasoning which led to the requirement of homogeneity and isotropy. Thus in our simple scenario, the particle ($r\leq r_0$) will be characterised by a constant density $K_{mat}$, which we will require to be positive so that small spheres have a larger volume to surface ratio than they do in a spacelike hypersurface in Minkowski space. This gives us a simple, spherical metric inside the particle,
\beq\label{eq:g inside particle}
\g |_{r\leq r_0} =
\begin{pmatrix}
0 & 0 & 0 & 0 \\
0 & \frac{1}{1-r^2 K_{mat}/3 } & 0 & 0 \\
0 & 0 & r^2 & 0 \\
0 & 0 & 0 & r^2 sin^2(\th)
\end{pmatrix}.
\eeq
Note that this implicitly assumes $r_0<\sqrt{3/K_{mat}}$.

The assumed symmetries mean that in both spacetime and matter regions the metric has the general form,
\beq\label{eq:g stat sph symm}
\g =
\begin{pmatrix}
\gtt & 0 & 0 & 0 \\
0 & \grr & 0 & 0 \\
0 & 0 & r^2 & 0 \\
0 & 0 & 0 & r^2 sin^2(\th)
\end{pmatrix},
\eeq
with two degrees of freedom, $\gtt$ and $\grr$. Following standard notation we will use the symbols `$r$' and `$t$' interchangeably with `$1$' and `$0$' in the indices of tensor components, so that $\gtt=g_{00}$, $R_{1010}=R_{rtrt}$ and so on. Then inside matter we have,
\bea
\gtt &=& 0 \label{eq:gtt inside matter} \\
\grr &=& \frac{1}{1-r^2 K_{mat}/3 }. \label{eq:grr inside matter}
\eea
In general we have,
\beq\label{eq:grr in terms of K23}
\grr = \frac{1}{1-r^2 K_{23}(r)},
\eeq
and,
\bea
\KS &=& 3\Ktf+r K_{23}'(r) \label{eq:KS K23 diff eqn} \\
\Rightarrow \Ktf &=& \frac{A}{r^3}+\frac{1}{r^3}\int_0^r r^2 \KS dr, \label{eq:K23 general integral KS with A}
\eea
where $A$ is a constant of integration. However comparing (\ref{eq:grr in terms of K23}) and (\ref{eq:grr inside matter}) we see that $A=0$ so that,
\beq\label{eq:K23 general integral KS}
\Ktf = \frac{1}{r^3}\int_0^r r^2 \KS dr.
\eeq
Notice that $\grr$ can be found from $\ks$ and vice versa, so where convenient we can think of $\ks$ as one of our two degrees of freedom in place of $\grr$.

We smoothly connect matter and spacetime regions by use of matter indicator functions, which will be constrained by our asymptotic requirement that $\g\rightarrow g_{Sch}$, which implies $\grr^{1/2}(-\gtt)^{1/2}\rightarrow 1$, as $r\rightarrow \infty$. It will be convenient to insert our matter indicator functions into $\ar = \grr^{1/2}(-\gtt)^{1/2}$ and $\KS$ rather than $\gtt$ and $\grr$, so we characterize the matter region by,
\bea
\ar &=& 0 \\
\KS &=& K_{mat}.
\eea
Asymptotically (as $r\rightarrow \infty$) we want,
\bea
\ar &\rightarrow& 1 \\
\KS &\rightarrow& 0.
\eea
So in general we will set,
\bea
\ar &=& \mt \label{eq:alpha equals mt}\\
\KS &=& K_{mat}(1-m_r(r)), \label{eq:KS in terms of mt}
\eea
where $\mt$ and $m_r(r)$ are matter indicator functions that coincide with the zero function in matter regions and are non-zero in spacetime. The unique analytic extension of $\mt|_{r\leq r_0}$ is the zero function, however we require $\mt$ to be non-zero for $r>r_0$, forcing it to be non-analytic at $r_0$. Similarly $m_r(r)$ will be non-analytic at $r_0$. For simplicity in our toy model we will set $m_r(r) = \mt$, though we do not expect this to hold in solutions of a plausible theory of gravity. The system is then reduced to one degree of freedom, described by $\mt$, in terms of which we can specify the metric and thus the whole geometry. Notice that in this coordinate system $\v=\d_t$ so that the usual matter indicator function is $(-\gtt)^{1/2}$; however noting that $(-\gtt)^{1/2}=\ordt{1}$ we find it more convenient to use $\mt$.

Turning to the asymptotic behaviour, we find it simpler to analyse if we switch from $r$ to $p=1/r$ and include the `infinite' point $p=0$ in our discussion. We then could formalise the notion that the metric `asymptotically behaves like the Schwarzschild metric' into the requirement that $\g(p)\rightarrow g_{Sch}(p)$ as $p\rightarrow 0$. However, since the Schwarzschild metric is itself asymptotically Minkowski, we would like to make this condition stronger to distinguish between convergence with the Schwarzschild and Minkowski metrics. A tidy means of achieving this would be to make $m_t(p)$ non-analytic at $p=0$, so that all the derivatives of $\g(p)$ converge with all the derivatives of $g_{Sch}(p)$ at $p=0$. However, as we are here primarily interested in exploring the `new' behaviour near matter rather than the `GR-like' asymptotic behaviour, we will for the purposes of this toy model be satisfied with a simpler and more tractable matter indicator function which is analytic at $p=0$, so that only its first few derivatives converge with those of $g_{Sch}$ as $p\rightarrow 0$,
$$
\mtp=1+p \mt'(p)|_{p=0}+p^2/2 \mt''(p)|_{p=0} +\ldots
$$
Then we can write,
\bea
\Ktfp &=& p^3 \int_{p}^{\infty} p^4 \KSp dp \nn \\
&=& Cp^3 -p^3 \int_{0}^{p} p^4 \KSp dp \nn \\
&=& Cp^3 +K_{mat} (p^5/2\mt'(p)|_{p=0} +p^6/3!\mt''(p)|_{p=0}+\ldots). \label{eq:K23p expansion}
\eea
Now in the Schwarzschild solution we have $\Ktfp=2Mp^3$ where $M$ is the central mass. Thus if we relate the constant of integration, $C=\int_0^{\infty}r^2\KS dr$ with the Schwarzschild mass by setting $C=2M$, (\ref{eq:K23p expansion}) recovers the Schwarzschild term with a higher order correction which we can tweak by requiring $(\d_p^{(i)}(\mtp)|_{p=0})=0$ whenever $i<n$, for some choice of $n$. Putting this back into the metric using (\ref{eq:grr in terms of K23}) and (\ref{eq:alpha equals mt}) yields,
\bea
\grrp &=& \frac{1}{1-2M p} + O(p^{n+2}) \label{eq:grr p expansion} \\
\gttp &=& -(1-2M p) + O(p^{n+2}). \label{eq:gtt p expansion}
\eea

\subsection{Curvature and Dynamics in the Toy Model}\label{subsec:curvature in toy model}

\subsubsection{The Matter Indicator Function}

We can now construct explicit toy models by specifying $\mt$, $r_0$ and $K_{mat}$. Some insight can be gained by comparing the behaviour of such a toy model with that of the Schwarzschild metric of the matter free solution to GR.  We will here explore the metric $\g$ given by,
\bea
r_0 &=& 2 \nn \\
K_{mat} &=& 1/8 \nn \\
\mt &=& e^{-(r-r_0)^{-6}}. \label{eq:mt toy model definition}
\eea
Asymptotically this yields,
\bea
\grrp &=& \frac{1}{1-2M p} + O(p^{8})  \\
\gttp &=& -(1-2M p) + O(p^{8}),
\eea
however we will be primarily focused on the behaviour near matter, so we will switch back to using $r$ rather than $p$. Then as we approach the particle, $r\rightarrow r_0$, we see that $\mt\rightarrow 0$ as expected. Furthermore, it is easy to see that as $r\rightarrow 0$ we have $(r-r_0)^{-i}\mt\rightarrow 0$ for all $i$, meaning that $f(r)\mt\rightarrow 0$ whenever there exists a positive integer $i$ such that $(r-r_0)^i f(r)$ is analytic, in which case we will say that $f(r)$ is \emph{Laurent (at $r_0$)}. Intuitively, the zero in $\mt$ at $r_0$ is `infinitely powerful' and overwhelms a pole of any degree. In fact, if $\mt$ were to be continued into the complex plane it would have an essential singularity at $r_0$.

\begin{figure}[htb]
\centering
\includegraphics[scale=0.7]{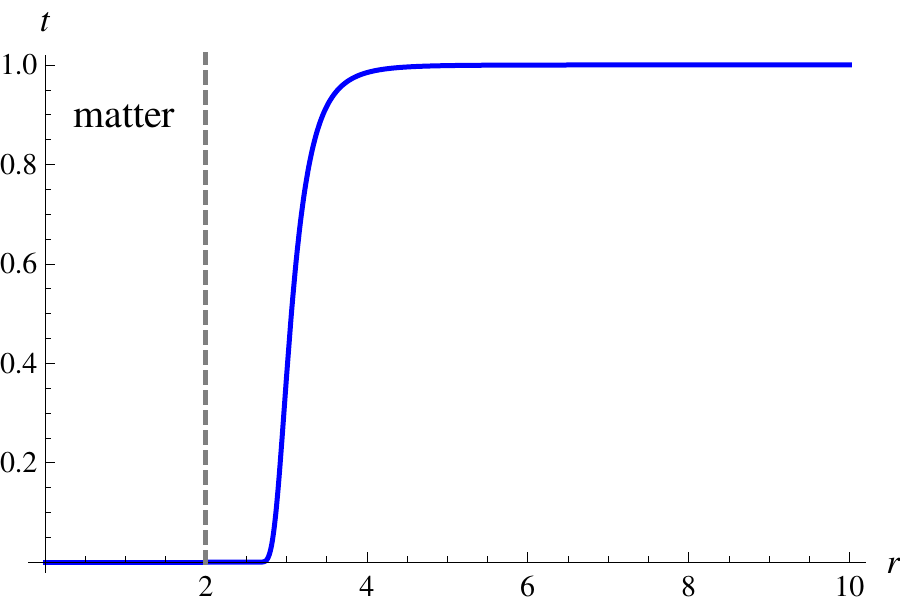}
\caption{The Matter Indicator Function $\mt$}
\label{fig:mt}
\end{figure}

This suggests that we can study the behaviour of geometric terms near matter by expanding them as a power series in $\mt$. We will say that a function $F(r)$ is order $i$ in $\mt$, or $F(r)=\ordt{i}$, if $F(r)/\mt^i$ is Laurent at $r_0$ (see the discussion of order in appendix \ref{appendix:lapG is order zero}). Now notice that $\mt'=6(r-r_0)^{-7}\mt$, so that $\mt^{(i)}=\mu_i(r)\mt$ where $\mu_i(r)$ is a polynomial in $(r-r_0)$ and thus Laurent, therefore $\mt^{(i)}=\ordt{1}$.

Because the matter indicator function is very close to $1$ asymptotically, and has an `infinitely powerful' zero at the boundary of matter, we will split spacetime into two `regions'; the \emph{asymptotic region} in which $\mt$ is not `noticeably' or `significantly' different from $1$, and the \emph{m-region} in which $\mt\neq 1$ becomes noticeable (and dominant). This split is not precise, depending on what exactly we mean by `noticeable', and we will not here attempt to make it so since we will use these regions for descriptive purposes only.

\subsubsection{Density}

Our assumed symmetries make the `$2$' and `$3$' indices interchangeable in many contexts, in particular we have $\krh=K_{13}$ (and $\kht=K_{03}$) so that $\ks=\khf+2\krh$. Since we have intuitively described $K_S$ as the `density' of space, we think of $\khf,\krh$ (and of course $K_{13}$) as the `space (sectional) curvatures'.

Now in the Schwarzschild metric we have,
\beq \nn
K_{Sch \, ab} =
\begin{pmatrix}
0 & -\frac{M}{r^3} & -\frac{M}{r^3} & \frac{2M}{r^3} \\
-\frac{M}{r^3} & 0 & \frac{2M}{r^3} & -\frac{M}{r^3} \\
-\frac{M}{r^3} & \frac{2M}{r^3} & 0 & -\frac{M}{r^3} \\
\frac{2M}{r^3} & -\frac{M}{r^3} & -\frac{M}{r^3} & 0
\end{pmatrix},
\eeq
so that the density is $K_S=0$. Thus recalling that the Schwarzschild metric is the unique solution to matter free GR given our assumed symmetries, $K_S=0$ takes us back to our intuition of density as a `mass field' which we expect to be zero valued in `empty' space.

Now in our toy model we have deliberately set $K_{mat}>0$ to represent the `massiveness' of the matter region, so that $K_S$ acts as a `mass field' inside our particle. However our requirement of smoothness in $\g$ means that the continuity of $K_S$ ensures that density is non-zero in spacetime, and though it falls `quickly' to zero it is `noticeably' positive in the m-region around the particle. Intuitively this means that our `mass field' extends into `empty spacetime'.

\begin{figure}[htb]
\centering
\includegraphics[scale=0.7]{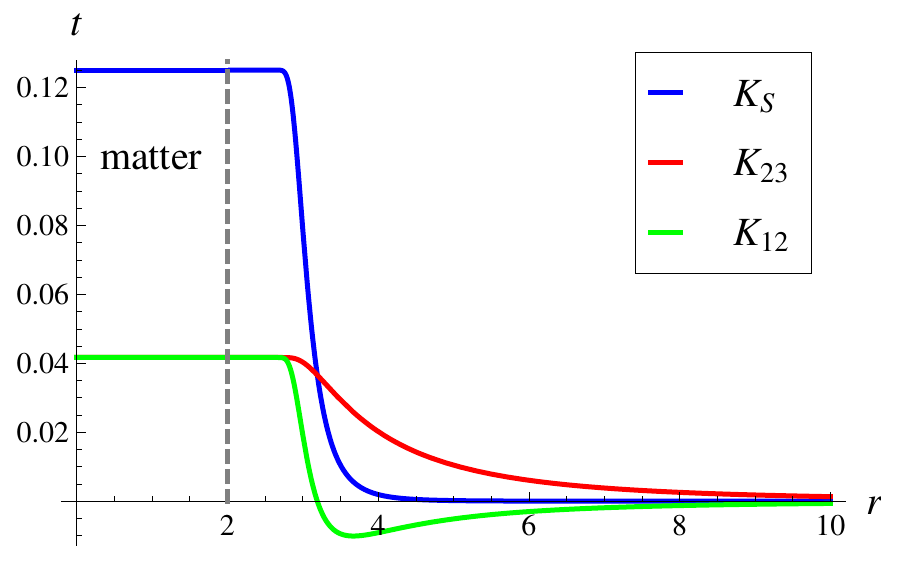}
\caption{Density}
\label{fig:density}
\end{figure}

As shown in figure \ref{fig:density}, $\khf$ and $\krh$ take the constant value $K_{mat}/3$ inside the particle, and then shift smoothly toward their asymptotic Schwarzschild behaviour, $\khf\rightarrow 2M/r^3$ and $\krh\rightarrow -M/r^3$.

Finally we progress to the Einstein tensor. Noting that in this coordinate system $\v=\d_t$ so that,
$$
G_{tt} = \v^a\v^bG_{ab},
$$
we can expect $G_{tt}$ to be well behaved near matter, and in fact we have,
$$
G_{tt} = -g_{tt}\ks.
$$
Thus as we approach matter the $g_{tt}$ term will force $G_{tt}\rightarrow 0$, and asymptotically $\ks\rightarrow 0$ will yield $G_{tt}\rightarrow 0$. However, $G_{tt}$ is `noticeably' non-zero in the m-region, in which we expect to see behaviour which differs from GR.
\begin{figure}[htb]
\centering
\includegraphics[scale=0.7]{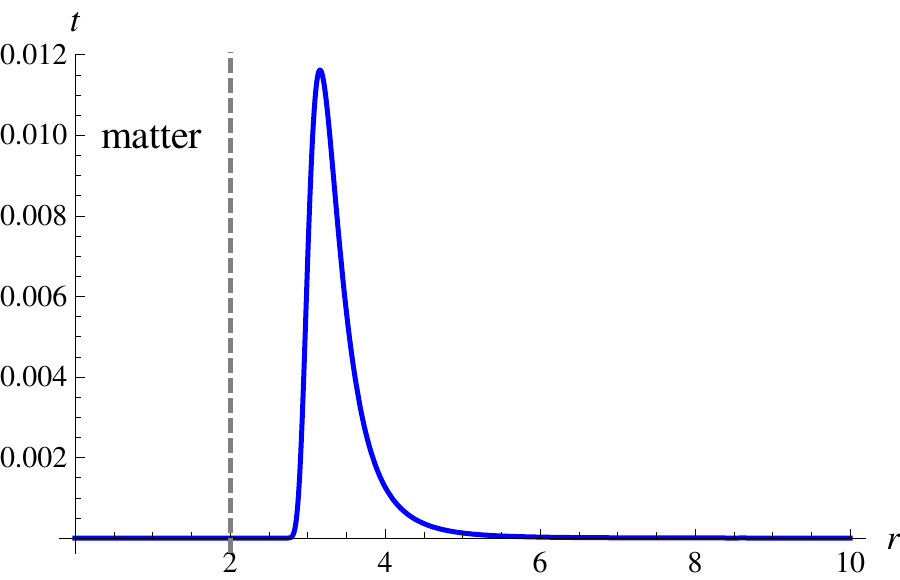}
\caption{The Einstein Tensor Component $G_{tt}$}
\label{fig:G44}
\end{figure}

\subsubsection{Acceleration}

\begin{figure}[htb]
\centering
\includegraphics[scale=0.7]{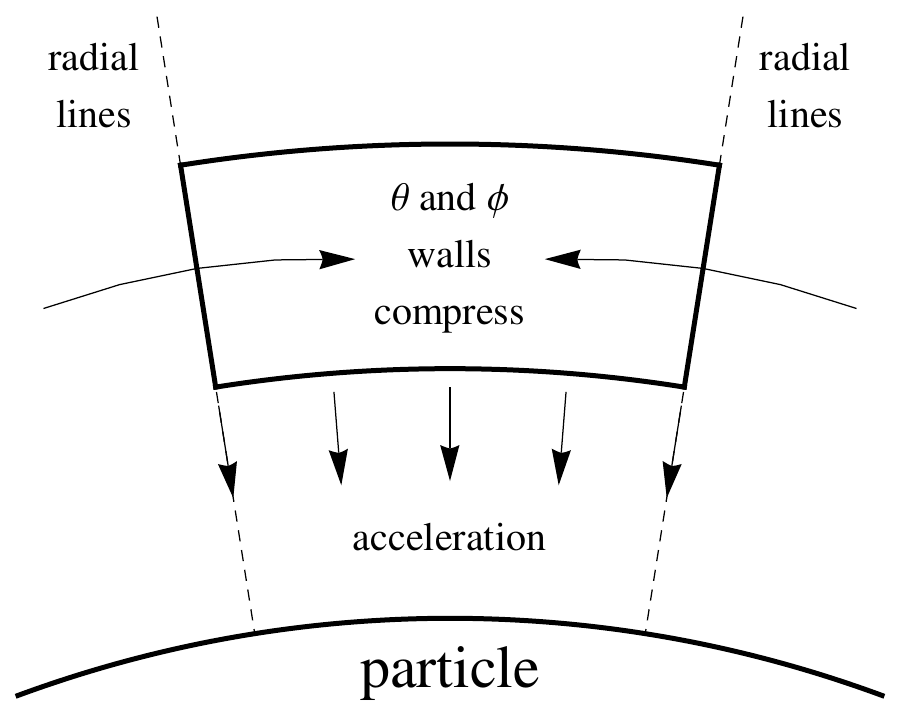}
\caption{The Infalling Box}
\label{fig:infallingbox}
\end{figure}

Just as we have interpreted $\ks$ in terms of density we seek to understand $K_t=\krt+2\kht$ (we think of $\krt$, $\kht$ as `time sectional curvatures') in terms of \emph{acceleration}. We start with an `infalling' point of space, released from rest (as perceived in this coordinate system) at the point $\{0,r,0,0\}$ to `fall' along a geodesic whose initial tangent vector at $t=0$ is $\d_t$, and ask what its `instantaneous acceleration' will be. This of course depends on how exactly we choose to measure `acceleration'; perhaps the most intuitive starting point being,
\bea
-\ddot{r} &=& \Gamma^r_{tt} \nn \\
&=& 2rg_{tt}\kht \nn \\
&=& \frac{2}{r}R_{2020}, \nn
\eea
where $\ddot{r}$ is the second derivative of $r$ with respect to proper time along the infalling point's geodesic path. It may at first seem counterintuitive that $\kht$ should be the curvature term corresponding to acceleration. To understand this better we advance our analysis from an infalling point to a small infalling coordinate `box', every point of which is travelling along a geodesic with initial tangent vector $\d_t$ at $t=0$. In the $t=0$ plane we can express the box as,
$$
[r-\delta_r,r+\delta_r]\times [-\delta_{\th},\delta_{\th}]\times [-\delta_{\phi},\delta_{\phi}].
$$
 The geodesics representing the path of the box form a congruence which we can parameterise by proper time. The symmetries of the system dictate that the geodesic motion will be radial, so we can think of every point in our infalling box as travelling along a radial line in space (figure \ref{fig:infallingbox}).
But then the convergence of these radial lines (as we approach $r=0$) entirely determines the change in the distance between the two `$\th$ walls' (which is initially $2\delta_{\th}$), and so the radial acceleration of the box entirely determines the acceleration in this relative distance (with a similar result in the $\phi$ direction).Thinking of the infalling box as a geodesic congruence, we see that the relative acceleration between the two $\th$ walls can be measured by $\kht$ (and that of the two $\phi$ walls by $K_{03}=\kht$), so we see how $\kht$ can be used as a measure of radial acceleration from rest.
\begin{figure}[htb]
\centering
\includegraphics[scale=0.7]{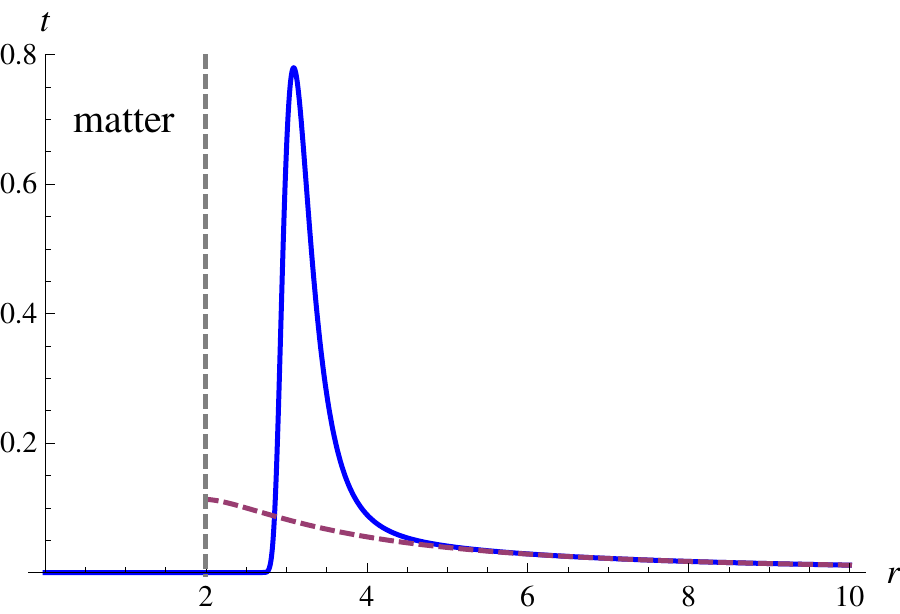}
\caption{Radial Acceleration $-\ddot{r}$}
\label{fig:negrdotdot}
\end{figure}
Figure \ref{fig:negrdotdot} shows $-\ddot{r}$ for both the Schwarzschild metric (dashed) and our toy model. Note that $\ddot{r}(r)$ represents the instantaneous acceleration of a point released from rest at $r$, \emph{not} the acceleration of a single infalling point. Thus $-\ddot{r}(r)$ can be thought of as describing a \emph{gravitational force field}. Notice that in the asymptotic region $-\ddot{r}_{toy}$ and $-\ddot{r}_{Sch}$ increase as $r$ decreases, following the classical result that gravity is `stronger' nearer the `source mass'. However this changes in the m-region, where $-\ddot{r}_{toy}$ deviates from $-\ddot{r}_{Sch}$ by spiking upward then falling to zero near the particle. This fall to zero occurs because of the smooth joining of spacetime and matter facilitated by $\mt$. As we approach matter spacetime begins to take on the characteristics of matter; since there is no time or change inside matter there can be no acceleration either.

\begin{figure}[htb]
\centering
\includegraphics[scale=0.7]{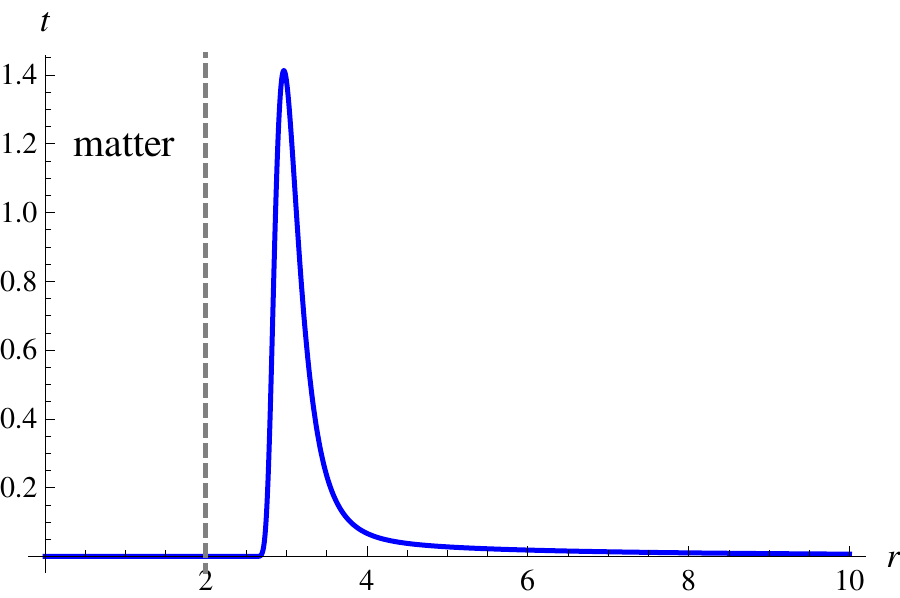}
\caption{Radial Acceleration $\thd$}
\label{fig:thetadot}
\end{figure}

\begin{figure}[htb]
\centering
\includegraphics[scale=0.7]{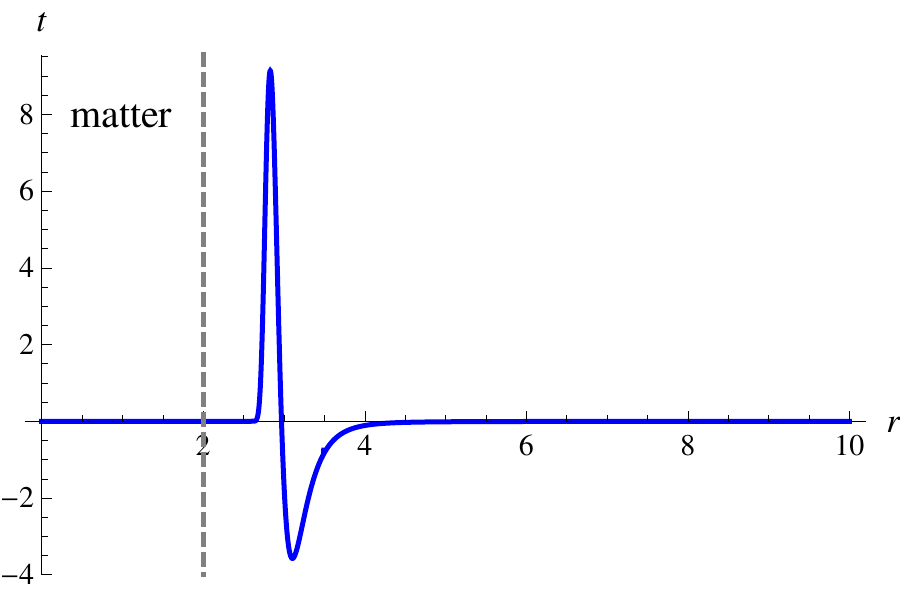}
\caption{Relative Radial Acceleration $\d_r\thd$}
\label{fig:dthetadot}
\end{figure}

Turning back to our small infalling coordinate box we have seen that the relative acceleration of the $\th$ (or $\phi$) walls is described by $\kht$, and how this is related to the radial acceleration. We now consider the relative acceleration of the $r$ walls, described by $\krt$, and expect this to be related to the $r$ derivative of $-\ddot{r}$ (since we are comparing the relative accelerations from rest of points at $r\pm \delta_r$). However the relationship is not simple; we have a congruence of geodesics with initial tangent vector $\d_t$ and initial points at $t=0$ running in the $r$ direction from $r-\delta_r$ to $r+\delta_r$. Connecting the points at proper time $s$ on these geodesics yields a hypersurface segment which can be thought of as the box after proper time $s$ according to an observer in the box. In particular, this surface will not in general lie entirely within a single $t=constant$ hypersurface.
\begin{figure}[htb]
\centering
\includegraphics[scale=0.7]{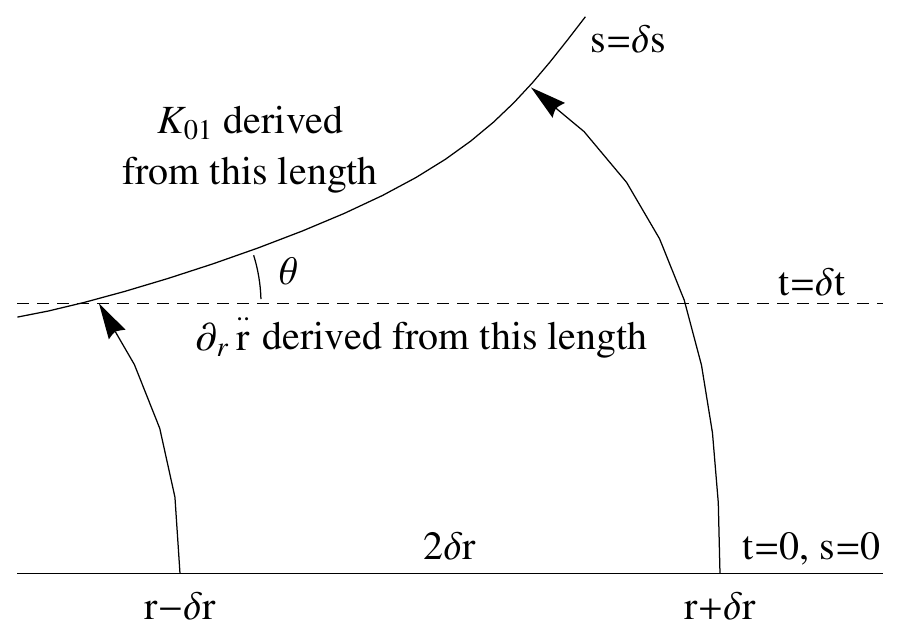}
\caption{$\d_r\ddot{r}$, $K_{01}$ and $\th$}
\label{fig:k14andthetadot}
\end{figure}
The calculation of $\krt$ compares $2\delta_r$ with the length of the image of this section of the r-axis in the `proper time $s$' hypersurface, whereas $-\d_r \ddot{r}$ will compare it to the length between the $r-\delta_r$ and $r+\delta_r$ geodesics along an r-axis in a $t=constant$ hypersurface; these two lengths are not in general the same (figure \ref{fig:k14andthetadot}). We label the angle between the proper time $s$ hypersurface and the $t=constant$ hypersurface by $\th(r,s)$, and notice that $\th$ itself can be used as an alternate measure of acceleration from rest. More precisely we can switch from $-\ddot{r}(r)$ to $\d_s\th(r,s)|_{s=0}$, which we will write simply as $\thd(r)$. It is easy to see that,
\bea
-\ddot{r}(r) &=& \frac{\sqrt{-\gtt}}{\sqrt{\grr}}\thd(r) \nn \\
\kht(r) &=& -\frac{1}{r\a(r)}\thd(r) \nn \\
\krt(r) &=& -\frac{1}{\a(r)}\d_r\thd(r).
\eea
While $\thd$ behaves like $-\ddot{r}$ (figure \ref{fig:thetadot}),
its radial derivative $\d_r \thd$ shows even more interesting behaviour (figure \ref{fig:dthetadot}). Thinking of $\d_r\thd$ as measuring the relative acceleration of the $r$ walls of our box, and thus its instantaneous `stretching' along the $r$ direction, we see that in the asymptotic region the box is stretched by our toy model's gravity just as it would be in the Schwarzschild metric. However as we traverse the m-region the overwhelming influence of $\mt$ decreases the acceleration so that the back end of the box accelerates faster than the front end, and the box contracts in the $r$ direction.

\begin{figure}[htb]
\centering
\includegraphics[scale=0.7]{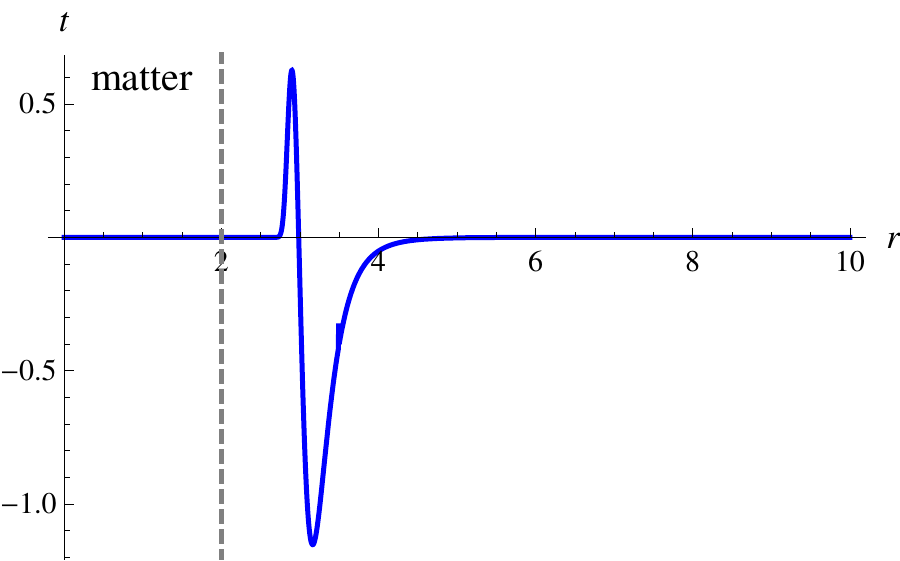}
\caption{Ricci Tensor Component $R_{tt}$}
\label{fig:R44}
\end{figure}

To see how this all affects the volume of the box we turn to the Ricci tensor, since $R_{tt}$ is typically thought of as a measure of the acceleration from rest of the volume of a small spatial three sphere along the geodesic congruence beginning at the sphere (at $t=0$) with initial tangent vector $\d_t$. The volume acceleration of a small sphere is equivalent to that of a small box, and while a sphere is perhaps more elegant a coordinate box is perhaps simpler; we will use both interchangeably. We have,
\bea
R_{tt} &=& \gtt K_t \nn \\
&=& \gtt(\krt+2\kht) \nn \\
&=& \frac{\sqrt{-\gtt}}{\sqrt{\grr}}(\d_r\thd+\frac{2}{r}\thd). \label{eq:R00 in terms of theta dot}
\eea
Figure \ref{fig:R44} shows the behaviour of $R_{tt}$. In the asymptotic region we see that the box gains in volume; unlike the Schwarzschild solution the $r$ axis stretching does not perfectly cancel out the $\th$ and $\phi$ contraction. Entering the m-region the volume acceleration becomes negative, with the $r$ axis contraction combining with the continued $\th$ and $\phi$ axes contraction to `squash' the box. In between these two behaviours is a `point' or more accurately `radius of harmony' at which the volume acceleration is zero. Finally, as we approach the boundary of matter the volume acceleration falls to zero as spacetime adopts the unchanging nature of matter.

\begin{figure}[htb]
\centering
\includegraphics[scale=0.7]{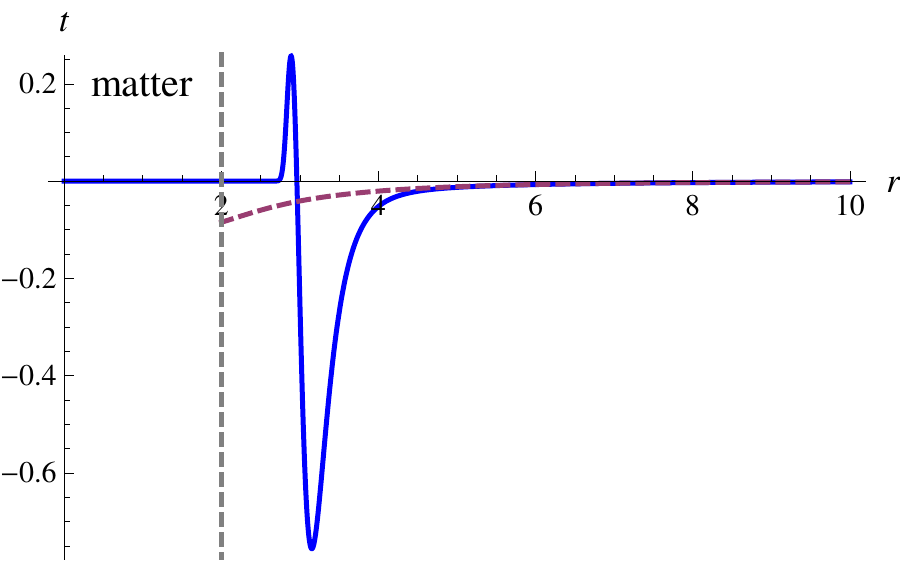}
\caption{Ovalness $Ov_{tt}$}
\label{fig:Ov}
\end{figure}

Because $R_{ab}=0$ in matter free GR the geometry must be characterised by the Weyl tensor $W_{abcd}$, which is intuitively $R_{abcd}$ with all the algebraic dependency on $R_{ab}$ (and thus $R$) subtracted out so that the remainder is entirely trace free. Following the treatment of the Riemann tensor, we can in this system capture all the degrees of freedom of the Weyl tensor in the `sectional Weyl curvatures',
\beq\nn
WK_{ij} = \frac{W_{ijij}}{g_{ii}g_{jj}-g_{ij}^2},
\eeq
which will typically be simpler than the components of the full Weyl tensor. In General Relativity the Weyl is usually thought of as describing the effects of mass at a distance, and so determined by boundary conditions (thus it can be non-zero in the matter free theory). Intuitively it is pictured as describing the volume change independent `stretching' or `changing in shape' of a small sphere (or coordinate box) accelerating from rest. To find a precise measure of the `change in shape independent of change in volume' of the small coordinate box in our toy model, to use alongside $R_{tt}$, we first note that our measure of volume acceleration is a sum of the time sectional curvatures\footnote{Notice that only in four dimensions is the number of space sectional curvatures (contributing to $G_{00}$) equal to the number of the time sectional curvatures (contributing to $R_{00}$).}, $R_{tt}=\gtt(\krt+2\kht)$. This suggests that the `stretching' of the coordinate box (or the `ovalness' of the small spheroid) might be measured by a difference in sectional curvatures, for example $\krt-\kht$. However these sectional curvature terms still contain some algebraic dependency on volume acceleration (though not necessarily in the $t$ direction), stripping this out gives us our measure of `ovalness',
\beq\label{eq:ovalness definition}
Ov_{tt} = \gtt(WK_{01}-WK_{02}).
\eeq
We show ovalness in figure \ref{fig:Ov}, and the stretching and contracting described in the discussions above are evident.

To summarise, in the asymptotic region a small sphere released from rest accelerates toward the particle and in so doing is stretched along the radial direction and contracts along the axial directions, developing into an oval spheroid which is longer along the radial direction than along the axial directions, as we would see in the Schwarzschild metric. However unlike Schwarzschild, in our toy model the Ricci tensor is non-zero so that the sphere gains volume as it stretches to become a spheroid. Moving into the m-region, the powerful influence of $\mt$ overwhelms the `gravity gradient' to cause the acceleration at the front end of the sphere to be less than the acceleration at the back; thus the sphere contracts along the $r$ direction while continuing as before to contract along the axial directions. Thus the `oval development' of the sphere reverses, and it becomes longer in the axial directions than in the radial direction. The sphere will to some degree `wrap' itself around the particle as the front and back ends take on the shape of the particle's surface. In this process the volume of the sphere decreases, and we identify a `radius of harmony' in between these two Ricci behaviours at which volume acceleration is zero. Finally as we approach the boundary of matter all of these effects disappear as spacetime smoothly adopts the unchanging nature of matter.

\subsection{The Equation}\label{subsec:equation}

In keeping with the physics tradition of thinking of `space variables' as explaining (or causing) `evolution in time', we now seek a connection between the space curvatures and the time curvatures as a path toward a new governing equation. Intuitively we are looking for density to `explain' acceleration in some way.

Our starting point is in noticing that the graph of $\kht$, or alternatively of $\thd$ or $-\ddot{r}$, looks as though it were a derivative of $\ks$. Intuitively this relates acceleration from rest (which we might think of as a manifestation of the `force' of gravity) with the slope of the density of space, so that a point of space will `roll uphill' along the density gradient toward the `peak' at the particle, which is generating the density slope. We will build on this insight to develop a governing equation. Our starting point is,
\beq
-\kht = \ks'.
\eeq
Now because of the way in which it neatly characterizes $\kht$, $\krt$ and thus $R_{tt}$, we will use $\thd$ in place of $-\kht$. Further, as we seek to end up with a tensor equation we will use curvature tensor components wherever possible rather than sectional curvatures. In particular we will be biased toward the Ricci and Einstein tensors predominant in GR. Then noting that $\nabla_r G_{tt}=-\gtt\ks'$, and introducing a scalar function $A(r)\in\R$ for generality, our next step is,
\beq\label{eq:thd equals AdelG}
-\thd = A(r) \nabla_r G_{tt}.
\eeq
Now in (\ref{eq:thd equals AdelG}) our measure of acceleration $\thd$ depends only upon the nature of the density at (or in a small neighbourhood of) that point, and is not directly influenced by density at a distance. This is in contrast with matter free GR, in which the density is zero in `empty' spacetime and acceleration is entirely due to the boundary effects of matter at a distance as expressed in the Weyl curvature. We are uncomfortable with this discrepancy as we would like our theory to behave like GR in the ranges in which the later has been successfully tested. We therefore introduce an additional (not necessarily scalar) term $B(r)$ which we expect to account for the effect of density at a distance so that while $\nabla_r G_{tt}$ may be dominant in the m-region, as $G_{tt}\rightarrow 0$ in the asymptotic region the $B_r(r)$ term will become dominant yielding our desired asymptotic Schwarzschild behaviour. Thus while the `$B$' or asymptotic gravity will be a slight adjustment of previous matter free GR, the $\nabla_r G_{tt}$ or m-region gravity is entirely new. We now have,
\beq\label{eq:thd=AdelG+B}
-\thd = A(r) \nabla_r G_{tt} + B_r(r).
\eeq

So far our equation has been scenario (and coordinate system) specific; we now seek a more general equation. As it stands, even if (\ref{eq:thd=AdelG+B}) were linear we would require a pre-knowledge of the position of every region of matter (with density $>0$) in the universe before being able to superimpose the various versions of (\ref{eq:thd=AdelG+B}) to yield a general equation. We wish to be able to express our equation independently of its various solutions.

To begin with, since it represents the effect of matter at a distance $B_r(r)$ may explicitly depend on the distance to the boundary of matter. To resolve this issue, recall that $\thd$ represents the acceleration of a point from rest; if instead we were to consider the relative accelerations of two nearby, radially separated points we might hope that our dependence on the distance to the boundary of matter will be `cancelled out'. Our differential equation would then have `moved up a degree', focusing on the curvature of the slope of the density rather than in its gradient. We then have,
\beq
-\d_r\thd = \widetilde{A}\nabla_r\nabla_r G_{tt} +\widetilde{B}_r.
\eeq
As previously stated we wish to end up with a tensor equation, so we move to,
\beq
R_{1010} = \nabla_1\nabla_1 G_{00} +\widetilde{B}_1.
\eeq
Next, we seek to eliminate the directional dependence so that we do not have to depend on pre-knowledge of the direction toward all massive matter, or explicitly include every such `gravitational source term' in the general equation. To achieve this we begin by summing over the spatial directions,
\beq
-\sum_{i\in\{1,2,3\}}R_{i0i0}=\sum_{i\in\{1,2,3\}}\nabla_i\nabla_i G_{00} +\sum_{i\in\{1,2,3\}}\widetilde{B}_i.
\eeq
This reminds us of our coordinate box (or small sphere), and the intuition is of a small spatial sphere `sitting' on the slope of the density distribution (the graph of $\ks(r)$), with each point of the sphere evolving along the normal to the slope, so that the curvature of the density slope determines the evolution of the shape of the sphere. This recalls our description of the change in the shape of a small sphere using the Ricci and Weyl tensors; we notice that the Ricci might relate to the Riemann term $-\sum_{i\in\{1,2,3\}}R_{i0i0}$ while the Weyl tensor might relate to the `effect of matter at a distance' term $\sum_{i\in\{1,2,3\}}\widetilde{B}_i$. Now $\sum_{i\in\{1,2,3\}}\widetilde{B}_i$ is unknown, and its exact form may still be solution dependent. Further, we still do not have a tensor equation. We can resolve these problems by implementing a full trace rather than the spatial sum above. We will assume that the Weyl-like terms $\widetilde{B}_i$ will cancel out leaving us with,
\beq
-R_{00}=\triangle G_{00}.
\eeq
Finally, to make this a tensor equation which is well behaved near matter (we discuss the order of the Laplacian term in section \ref{appendix:lapG is order zero}), we will return to a three observer framework and in the intersection basis write,
\beq\label{eq:Governing Equation of Gravity}
R_{AB} + \triangle G_{AB} = 0.
\eeq
Notice that the matter free Einstein equation is automatically a solution.

We now have a first attempt at a \emph{governing equation for a complete theory of gravity}, which we hope is the first step toward an ultimate `\emph{underlying equation}' describing all observed particles, charges, fields and forces.

\section{Summary and Discussion}\label{sec:conclusionCG}

\subsection{Summary}\label{subsec:summaryCG}

In this chapter we take the first steps toward a complete theory of gravity in the Matter-Space-Time framework, developing a governing equation by consideration of a static, spherically symmetric `Schwarzschild' scenario  with a central stationary spherical particle. We reject the matter free Einstein Equation on the grounds that it can not smoothly join matter and spacetime regions, instead building a simple toy model based on representing the `mass field' with the total curvature of the particle's three geometry. We use the insight gained from this toy model to propose a new governing equation for a complete theory of gravity.

\subsection{Looking Ahead}\label{subsec:looking aheadCG}

Looking ahead to the further development of this theory we briefly discuss some areas which may be of immediate interest.

\bd
\item[\emph{The One-Body Problem:}] We could seek to solve the governing equation (\ref{eq:Governing Equation of Gravity}) in the simple `Schwarzschild' scenario used in our toy model. In essence this would act as a `proof of concept' for our equation, demonstrating that our theory allows solutions which smoothly join matter and spacetime.
\item[\emph{The Two-Body Problem:}] We could seek to solve the governing equation in the simplest possible scenario containing two particles initially at rest relative to one another. The particles could be identical, or one might be much more massive than the other. We anticipate that such solutions would take us a long way toward understanding how the `force' of gravity behaves in our theory.
\item[\emph{Linearisation:}] Suitably linearising the equation would allow us to make first order statements about large numbers of particles.
\item[\emph{Cosmology:}] Finding solutions to our equation which would take the place of the FRW solution in GR might allow us to examine cosmology and large scale structure in our theory.
\item[\emph{The Consequences of Non-Zero Ricci Curvature:}] Unlike GR our theory allows non-zero Ricci curvature in `empty' spacetime. This may have implications for the calculation of astronomical distances or even the `apparent' speed of light (as measured using distances calculated `incorrectly' assuming zero Ricci curvature).
\item[\emph{Black Holes:}] Our new equation may lead to interesting new behaviour in geometries involving an event horizon.
\item[\emph{Compatibility with General Relativity:}] We would of course like our theory to be compatible with GR at the scales at which it has been successfully tested, and will have to check that this is the case.
\item[\emph{Differences with General Relativity:}] We would like to find and investigate the differences between our theory and GR, since such differences could potentially lead to falsifiable predictions and experimental verification.
\item[\emph{Further Development of the Governing Equation:}] Our equation should be regarded as a first step; it has been developed based on the simplest of scenarios so that consideration of more complicated scenarios may lead to its modification, as might experimental analysis.
\item[\emph{Other Forces:}] We hope that the MST framework is flexible enough to accommodate other forces, which would take more complicated particle three geometries as their `sources'. We anticipate that our governing equation would have to undergo modification to deal with more forces; nevertheless the most natural starting point would be to explore solutions to our existing equation using more complicated matter three geometries.
\ed

\chapter[Toward a Schwarzschild Solution]{Toward a Schwarzschild Solution}

\begin{small}
\begin{center}
   \textbf{Abstract}
  \end{center}
   In this chapter we take the first steps toward solving the governing equation of gravity suggested in chapter $2$ in a static, spherically symmetric MST manifold. By exploring various expressions of the governing equation we are able to reduce it in this system to a pair of equations which are first order in the curvature.
\end{small}
\vskip 0.5cm

\section{The Governing Equation in a Static, Spherically Symmetric System}

\subsection{Various Forms of the Governing Equations}

In general we have,
\beq
R_{AB}+\lap G_{AB} = 0.
\eeq
Outside matter, in a single observer's coordinate system we can use the usual forms of our tensors to give,
\beq\label{eq:governing equation in chap 3}
R_{ab}+\lap G_{ab} = 0.
\eeq
Taking the trace yields,
\beq\label{eq:lapR=R}
\lap R=R.
\eeq

In the static, spherically symmetric system described in the last chapter only the diagonal terms of the Ricci and Einstein tensors are non-zero, and the $\{33\}$ term is a multiple of the $\{22\}$ term. We are left with three equations,
\bea
R_{00}+\lap G_{00} &=& 0 \nn \\
R_{11}+\lap G_{11} &=& 0 \nn \\
R_{22}+\lap G_{22} &=& 0. \label{eq:Rii plus lapGii}
\eea
Note that in this coordinate system $\v_X=\d_0$ while the $\d_i$ are spacelike. Thus the $\{00\}$ equation stands out since we expect $R_{00}$ and $\lap G_{00}$ to remain finite as we approach matter. Notice also that we have three equations and two degrees of freedom so a solution is not guaranteed. However it is easy to see that solutions to $R_{ab}=0$ will satisfy the above, so that in the Schwarzschild metric is a solution to (\ref{eq:Rii plus lapGii}), corresponding to a matter free universe. We will wish to explore more general solutions, and in particular to find solutions including regions of matter.

These equations simplify if we take simple combinations. Setting,
\beq\label{eq:def of En}
F_{ab} = \frac{G_{aa}}{g_{aa}}-\frac{G_{bb}}{g_{bb}},
\eeq
and taking the related combinations of (\ref{eq:Rii plus lapGii}),
\beq
g_{rr}\left(\frac{1}{g_{aa}}(R_{aa}+\lap G_{aa})-\frac{1}{g_{bb}}(R_{bb}+\lap G_{bb})\right),
\eeq
we get a second form of the governing equations, this time involving two equations,
\bea
F_{14}''+D_r F_{14}'+g_{rr}F_{14} &=& \left(\frac{g_{tt}'}{g_{tt}}\right)^2F_{14}+\left(\frac{2}{r}\right)^2F_{12} \nn \\
F_{12}''+D_r F_{12}'+g_{rr}F_{12} &=& \frac{1}{2}\left(\frac{g_{tt}'}{g_{tt}}\right)^2F_{14}+\frac{3}{2}\left(\frac{2}{r}\right)^2F_{12}, \label{eq:En}
\eea
where we use the dash notation to denote an $r$ derivative, $f'=\d_r f$, and where,
\beq
D_r = \frac{1}{2}\left(\frac{g_{tt}'}{g_{tt}}-\frac{g_{rr}'}{g_{rr}}+\frac{4}{r}\right).
\eeq
Finally, (\ref{eq:lapR=R}) can be written in this system as,
\beq\label{eq:R diff equation}
R'' + D_r R' -\grr R = 0.
\eeq

\subsection{The Bianchi Identities}

We can express a Bianchi identity in terms of the $F_{ab}$. In four dimensions the Bianchi identities are equivalent to,
\beq
W^a_{bcd;a} = R_{b[d;c]}-\frac{1}{6}g_{b[d}R_{;c]}.
\eeq
Now note that,
\bea
W^a_{414;a} &=& \frac{g_{tt}}{r^3}(r^3 WK_{14})' \nn \\
W^a_{212;a} &=& \frac{1}{r}(r^3 WK_{24})', \nn
\eea
so that,
\bea
\left(\frac{r^3 Ov_{tt}}{g_{tt}}\right)' &=& \left[r^3(WK_{14}-WK_{24})\right]' \nn \\
&=& \frac{2r^2 g_{tt}^2 - r^3 g_{tt} g_{tt}'}{4g_{rr}g_{tt}^2}R_{11} + \frac{r^3}{2\gtt}R_{44}'-\frac{r^3\gtt'}{4\gtt^2}R_{44}-\frac{r}{2}R_{22}'+\frac{1}{2}R_{22} \nn \\
&=&\frac{r^3}{2}\left((F_{12}'+\frac{1}{r}F_{12})-(F_{14}'+\frac{1}{2}\frac{\gtt'}{\gtt}F_{14})\right). \nn
\eea
Now note that $WK_{ij}$ has only one degree of freedom given our assumed symmetries, namely,
\beq
W = WK_{24} = \frac{1}{6}(\krh-\krt-\khf+\kht).
\eeq
Then since $WK_{14}=-2W$ we have $Ov_{tt}=-3\gtt W$, and we can write our Bianchi identity as,
\beq\label{eq:En Bianchi}
\frac{6(r^3W)'}{r^{3}}-\frac{(\sqrt{-\gtt}F_{14})'}{\sqrt{-\gtt}}+\frac{(r F_{12})'}{r} = 0.
\eeq

\section{Reducing the Equations to First Order}\label{sec:reducing the governing equation to first order}

It may not be possible to find a general `closed form' solution to (\ref{eq:governing equation in chap 3}) in this system, even if particular closed form solutions can be found upon assumption of boundary conditions. We can however reduce the order of the equations in the general static, spherically symmetric case, and will do so for each of the `forms' of our equations, (\ref{eq:Rii plus lapGii}), (\ref{eq:En}) and (\ref{eq:R diff equation}) outlined above.

\subsection{Reducing the $R_{00}$ equation}\label{subsec:reducing R00+lapG00}

Of the three independent equations (\ref{eq:Rii plus lapGii}) the $\{00\}$ equation,
\beq\label{eq:R00+lapG00=0}
R_{00}+\lap G_{00} = 0,
\eeq
stands out for two reasons. Firstly in this coordinate system $\v=\d_0$ so that the tensors in (\ref{eq:R00+lapG00=0}) will be finite as we approach matter. Secondly, looking back to section \ref{subsec:equation} recall that we arrived at the governing equation (\ref{eq:Governing Equation of Gravity}) through an intuitive argument assuming (\ref{eq:R00+lapG00=0}) and its first order counterpart (\ref{eq:thd=AdelG+B}) in this static, spherically symmetric system. We therefore focus on (\ref{eq:R00+lapG00=0}), and in particular would like to derive a concrete realisation of (\ref{eq:thd=AdelG+B}).

To begin with, we can decompose the Laplacian sum,
\bea
\nabla_0 \nabla_0 G_{00} &=& -\G^1_{00}(\nabla_1 G_{00}) -2\G^1_{00}(\nabla_0 G_{01}) \nn \\
\nabla_1 \nabla_1 G_{00} &=& (\nabla_1 G_{00})' - (\G^1_{11}+2\G^0_{10})(\nabla_1 G_{00}) \nn \\
\nabla_2 \nabla_2 G_{00} &=& -\G^1_{22}(\nabla_1 G_{00}) \nn \\
\nabla_3 \nabla_3 G_{00} &=& -\G^1_{33}(\nabla_1 G_{00}), \label{eq:lapG00 decomposition}
\eea
which gives us,
\bea
\lap G_{00} &=& \frac{1}{\grr}(\nabla_1 G_{00})' \nn \\
&& -\left(2\frac{\G^0_{01}}{\grr}+\frac{\G^1_{00}}{\gtt}+\frac{\G^1_{11}}{\grr}+\frac{\G^1_{22}}{r^2}+\frac{\G^1_{33}}{r^2 sin^2(\th)}\right)(\nabla_1 G_{00}) \nn \\
&& -2\frac{\G^1_{00}}{\gtt} (\nabla_0 G_{01}), \nn
\eea
leading to,
\beq \nn
\grr \lap G_{00} = (\nabla_1 G_{00})' + C_r (\nabla_1 G_{00}) +\frac{\gtt'^2}{2\gtt} F_{14},
\eeq
where,
\beq \nn
C_r = \frac{1}{2}\left(-\frac{g_{tt}'}{g_{tt}}-\frac{g_{rr}'}{g_{rr}}+\frac{4}{r}\right).
\eeq
Then we can write (\ref{eq:R00+lapG00=0}) as,
\beq\label{eq:dG00 in terms of R00}
(\nabla_1 G_{00})' + C_r (\dGttr) = -\grr \Rtt - \frac{\gtt'^2}{2\gtt} F_{14}.
\eeq
Now setting,
\bea
\b_1 &=& \frac{r^2 \gtt'^2}{2 \a \gtt}F_{14} \nn \\
\b &=& -\frac{\a}{r^2} \int \b_1 dr \nn,
\eea
and using (\ref{eq:R00 in terms of theta dot}) we see that $\dGttr=-\a\thd$ solves $LHS = \grr\Rtt$ whereas $\dGttr=\b$ solves $LHS=- \gtt'^2/(2\gtt) F_{14}$. Then the linearity of (\ref{eq:dG00 in terms of R00}) means that the sum is a solution,
\beq\label{eq:little eqn}
-\a \thd = \dGttr + \frac{\a}{r^2} \int \b_1 dr.
\eeq
Recalling the intuitive argument of section \ref{subsec:equation}, we see that (\ref{eq:little eqn}) is the concrete realisation of (\ref{eq:thd=AdelG+B}) which we desired.

As discussed in section \ref{subsec:equation} the intuition behind (\ref{eq:little eqn}) is of a point `rolling uphill', accelerating up the slope of the density $G_{00}(r)$. Given the discussion of section \ref{subsec:equation} we see that the $\thd$ term on the $LHS$ represents the acceleration of an infalling point while the $\nabla_1G_{00}$ term on the $RHS$ represents the slope of the density. The integral term on the $RHS$ represents the contribution to the density slope as perceived by an infalling point by the effect of the change in frame (via parallel translation) along the infalling geodesic. To make this more clear we look back to (\ref{eq:lapG00 decomposition}) and note that the integrand $\b_1$ derives from the $\G^1_{00}(\nabla_0 G_{01})$ term in $\nabla_0\nabla_0G_{00}$. Notice that this is the only term in the decomposition of the Laplacian which is not expressed in terms of $\nabla_1G_{00}$. Now we expect a point released from rest at $t=0$ to accelerate toward $r=0$ with no axial motion (due to the symmetries of the system), in other words geodesics with initial tangent vector $\d_0$ at $t=0$ will remain in the $(r,t)$ plane and `curve' toward $r=0$. Then the basis vectors $\d_a$ at $t=0$ will be parallel transported to $T_s(\d_a)$, with the symmetries of the system ensuring that $T_s(\d_0)$ is a linear combination of $\d_0$ and $\d_1$. Intuitively, in this new frame an infalling point would measure spatial density to be $G(T_s(\d_0),T_s(\d_0))$ rather than $G_{00}$, and will therefore experience an acceleration in density due to the difference between the infalling and coordinate frames (this effect only appears at the second order and so contributes to acceleration). This acceleration is represented by the $\nabla_0G_{01}$ term, leading to the $\b_1$ term, which we integrate to yield a contribution to the density slope (first order) as seen by the infalling point.

\subsection{Reducing the $F_{ab}$ equation}\label{subsec:reducing the En eqns}

Turning to (\ref{eq:En}) we notice that the $F_{14}$ term on the $RHS$ is reminiscent of $\b_1$. We therefore take a combination of the $F_{14}$ and $F_{12}$ equations in which the $F_{12}$ term on the $RHS$ cancels out; we can then integrate and compare with (\ref{eq:little eqn}). We begin by defining,
\beq\nn
y=3F_{14}-2F_{12}.
\eeq
Then adding the $F_{14}$ and $F_{12}$ equations (\ref{eq:En}) with the appropriate weights we have,
\beq\nn
y'' + D_r y' = -\grr y +2\left(\frac{\gtt'}{\gtt}\right)^2 F_{14},
\eeq
leading to,
\beq\nn
\left(r^2\frac{\sqrt{-\gtt}}{\sqrt{\grr}} y'\right)' = -r^2 \a y - 4\b_1,
\eeq
so that,
\beq\nn
\frac{\a}{r^2}\int\b_1 dr = \frac{1}{4} \gtt y' -\frac{1}{4}\frac{\a}{r^2}\int r^2\a y \ dr.
\eeq
We can insert this into (\ref{eq:little eqn}). Now notice that $\nabla_1 G_{00}=-\gtt K_S'$ and that $y=(4K_S-R)$ so that $y$ is the difference between the spatial $K_S$ and spacetime $R$ densities (or alternatively a measure of the deviation of our system from a space of constant curvature). This leads to,
\beq\nn
\a\thd = \frac{1}{4}\gtt R' + \frac{1}{4}\frac{\a}{r^2} \int r^2 \a y \ dr.
\eeq
Finally, notice that we could alternatively express $y$ as $y=2(K_S-K_t)$ and that the integral of $r^2 \a K_t$ is $r^2 \thd$. We are left with,
\beq\label{eq:second little eqn}
6\a\thd = \gtt R' + 2\frac{\a}{r^2}\int r^2 \a K_S \ dr.
\eeq

\subsection{Reducing the $R$ equation}\label{subsec:reducing the R eqn}

We can simplify (\ref{eq:R diff equation}) by `normalising' the $r$-axis; switching from $r$ to,
\beq\nn
\r = \int_0^r \sqrt{\grr} \ dr,
\eeq
which parameterises the $r$-axis using metric length, so that,
\bea
\d_{\r} &=& \frac{1}{\sqrt{\grr}} \d_r \nn \\
g(\d_{\r},\d_{\r}) &=& 1. \nn
\eea
Then (\ref{eq:R diff equation}) can be written as,
\beq\label{eq:R diff eq in rho}
\d_{\r}\d_{\r}R+\d_{\r}(\ln(r^2\sqrt{-\gtt}))\d_{\r}R - R = 0.
\eeq
This leads to,
\beq\label{eq:little eqn R}
r^2\sqrt{-\gtt}\d_{\r}R = \int r^2 \sqrt{-\gtt} R \ d\r.
\eeq
Notice the similarity between the $RHS$ and the Einstein-Hilbert action, which we can find by multiplying by $\sin\th$ and integrating,
\beq\label{eq:EH action}
\int \sqrt{-g} \ \d_{\r} R \ dt d\th d\phi = \int R \sqrt{-g} \ d^4x.
\eeq

Alternatively we could write (\ref{eq:R diff eq in rho}) as,
\beq\nn
\d_{\r}(\ln(r^2\sqrt{-\gtt})) = \frac{R-\d_{\r}\d_{\r}R}{\d_{\r}R},
\eeq
leading to,
\beq\label{eq:R exp}
r^2\sqrt{-\gtt} = \frac{1}{\d_{\r}R}e^{\int \frac{R}{\d_{\r}R} d\r}.
\eeq

\section{Summary and Discussion}

\subsection{Summary}

In a static, spherically symmetric system we have derived three first order equations from the Governing Equation. Since a static, spherically symmetric metric has only two degrees of freedom we can discard one of these first order equations; we have then reduced the Governing Equation to,
\bea
\a \thd + \dGttr + \frac{\a}{r^2} \int \b_1 dr &=& 0\nn \\
r^2\sqrt{-\gtt}\d_{\r}R - \int r^2 \sqrt{-\gtt} R \ d\r &=& 0. \label{eq:little eqns}
\eea
Further, in this system we can use the Bianchi identities to yield an equation connecting Weyl and Ricci curvature,
\beq
\frac{6(r^3W)'}{r^{3}}-\frac{(\sqrt{-\gtt}F_{14})'}{\sqrt{-\gtt}}+\frac{(r F_{12})'}{r} = 0.
\eeq

\subsection{Looking Ahead}

Ideally we would be able to find a closed form general solution to (\ref{eq:little eqns}), however we might first have to specify the boundary conditions, further there may not be closed form solutions corresponding to all choices of boundary conditions. Given the symmetries of the system boundary conditions would consist of the radial position and the three geometry of matter, which is complicated by the fact that our manifold need not have $\R^4$ topology; in particular we would expect more a more complicated topology in some black hole solutions. Some examples of the possible matter distributions are:
\bd
\item{\bf No Matter:} The matter free Einstein equation is automatically a solution to the governing equation, however we would ideally like to find solutions which are singularity free.
\item{\bf No Spacetime:} A trivial solution in which the universe is entirely matterlike.
\item{\bf Central Particle:} A solution involving a particle centered at the origin. The particle may or may not lie inside an event horizon, and we may or may not have distant matter. If there there is no matter other than the central particle, we might impose Schwarzschild behavior on the metric as a further boundary condition.
\item{\bf Distant Matter:} Matter whenever $r\geq r_1$ for some $r_1>0$. We may or may not have a central particle.
\item{\bf Concentric Shells:} We could in theory have concentric shells of matter and spacetime regions, however the spacetime regions would not be able to interact and would not be causally related.
\ed

\appendix
\chapter{The Laplacian Near Matter}\label{appendix:lapG is order zero}

We will prove the following theorem.

\bT\label{thm:lapG is order 0}
$\triangle G_{AB}=\ordc{0}$.
\eT

We begin with a related lemma,

\bL\label{lemma:Gvv is order 2}
$G_{AB} = \ordc{2}.$
\eL
\begin{proof}
We start by noting that if we adopt an orthonormal basis in a spacetime region (so that the metric takes the Minkowski form), the diagonal components of the Einstein tensor are given by sums of sectional curvatures. Thus given a basis $\{e_0,e_1,e_2,e_3\}$ with $e_0$ timelike, $e_i$ spacelike (where $i\in\{1,2,3\}$), we have $G(e_0,e_0)=-(K_{12}+K_{13}+K_{23})$. Now $G(e_0,e_0)$ is independent of the choice of spatial basis, depending only on $G$ and $e_0$. Thus in our usual $3+1$ decomposition based on an observer $\Ob_X$ we conclude that $G(\n_X,\n_X) = -\ks$. But then $G(\v_X,\v_X)=-m_X^2 \ks$, and since $\ks$ is purely spatial and thus of order zero we see that $G(\v_X,\v_X)$ is of order $2$ in $m_X$. Then since $\v_B=\v_A/\v_A^{\overline{B}} + \orda{2}$, and since order is coordinate (and observer) independent we can conclude that $G(\v_A,\v_B)$ is order $2$ according to any observer.
\end{proof}

In what follows we will encounter and have to classify terms such as $\d_a (m_A)/m_A$ by their order in $m_A$. We have so far been assuming that we can expand the components of all geometric objects as series,
$$
F=\sum_i f_i m_A^i,
$$
where $f_i$ are Laurent functions of the coordinates; this may not be true in general. In particular, in our toy model (section \ref{subsec:curvature in toy model}) $\d_r (\mt)/\mt$ was Laurent so that $\d_r(\mt)$ was $\ordt{1}$. However, if for example we were to use,
$$
\mt=e^{-e^{(r-r_0)^{-2}}},
$$
then we would have $\d_r(\mt)/\mt=-2(r-r_0)^{-3}e^{(r-r_0)^{-2}}$ which is not Laurent. It seems natural to accommodate such possibilities by generalising our understanding of what we mean by a function being of order $m_A$; we will however leave this question to later research and for now continue to assume an expansion with Laurent coefficients and that $\d_a^{(i)}(m_A)/m_A$ is Laurent (so that $\d_a^{(i)}(m_A)=\orda{1}$). Note that to make this simple notion of order well defined we may have to restrict coordinate transformations to be analytic.

Turning to our theorem, we begin with an $MST$ manifold $\M$ and a three observer system $\{X,Y,Z\}$. Then any point $p\in U_{XYZ}$ lies in a spacelike hypersurface $\Sxp$ defined as the set of points in $U_X$ with the same coordinate time $x^0(p)$ as the point $p$, according to $\Ob_X$. Now the vector field $\v_X$ is orthogonal to $\Sxp$, and so by integrating $\v_X$ we can in some neighbourhood $U_{\{X,p\}}\subset U_X$ of $p$ find a coordinate system $\{v^0,v^1,v^2,v^3\}$ with $\d_{v^0}=\v_X$ and $\d_{v^i}$ spacelike and orthogonal to $\d_{v^0}$ (we will in what follows assume that the indices $i,j,k$ run from $1$ to $3$ and represent spacelike coordinate directions). We will remain with this $V$ coordinate system for the remainder of this section, writing $\d_a$ to mean $\d_{v^a}$ and so forth. Because we have previously shown (lemma \ref{lemma:m coord independence}) that order is coordinate independent, any such result we find in the $V$ system will have general application. Notice moreover that $\v_V=\d_0$ so that $-m_V^2=g(\d_0,\d_0)=g(\v_X,v_X)=-m_X^2$.

We can define $h_{Vij}$ to be the restriction of the metric to the spacelike hypersurfaces $\S_{\{V,v^0\}}$ of the $V$ coordinate system, with inverse matrix $h_V^{ij}$. Then the usual $3+1$ decomposition of the full metric yields,
\bea
g_{00} &=& -\mv^2 \nn \\
g_{0i} &=& 0 \nn \\
g_{ij} &=& h_{Vij} \nn \\
g^{00} &=& -\mv^{-2} \nn \\
g^{0i} &=& 0 \nn \\
g^{ij} &=& h_V^{ij}, \label{eq:g decomposition in V system}
\eea
from which we can see that the singular behaviour near matter is restricted to the single inverse metric component $g^{00}$. Thus when it comes to identifying the singular behaviour in our connection coefficients, we need only focus on the $\G^0_{bc}$ terms. Then requiring $i,j\in\{1,2,3\}$ to represent spatial coordinate directions as usual this gives us three cases, $\G^0_{ij}$, $\G^0_{i0}$ and $\G^0_{00}$, to which we will add a fourth case $\G^i_{00}$,
\ben
\item $\G^0_{ij} = 1/2\mv^{-2} g_{ij,0} = \ordv{-2}$. However, we could also write,
$$
\G^0_{ij} = 1/2 \mv^{-1} (\L_n g)_{ij},
$$
so the derivative assumption gives us $\G^0_{ij} = \ordv{-1}$.
\item $\G^0_{0i} = 1/2\mv^{-2} \d_i(\mv^2) = \ordv{0}$.
\item $\G^0_{00} = 1/2\mv^{-2} \d_0(\mv^2) = \ordv{0}$.
\item $\G^i_{00} = -1/2\sum_{j\in\{1,2,3\}}h_V^{ij}\d_j(\mv^{2}) = \ordv{2}$.
\een

We are now in a position to examine the covariant derivatives of $G_{00}$.
\bd
\item[1. $\nabla_a G_{00}$]
\beq\nn
\nabla_a G_{00} = \d_a(G_{00}) - 2\G^p_{a0}G_{p0}.
\eeq
Now,
 \begin{itemize}
      \item $G_{00}$ is order $2$ so by the above $\d_a(G_{00})$ is also order $2$.
      \item $G_{a0}$ is order $0$.
      \item $\G^p_{a0}$ is order $0$.
    \end{itemize}
Therefore $\nabla_a G_{00}$ is order $0$.

\item[2. $\nabla_0 G_{00}$]
\beq\nn
\nabla_a G_{00} = \d_0(G_{00}) - 2\G^0_{00}G_{00}-2\sum_{i\in\{1,2,3\}}\G^i_{00}G_{i0}.
\eeq
Now,
 \begin{itemize}
      \item $G_{00}$ is order $2$ so by the above $\d_0(G_{00})$ is also order $2$.
      \item $G_{00}$ is order $2$ and $\G^0_{00}$ order $0$.
      \item $G_{i0}$ is order $0$ and $\G^i_{00}$ order $2$.
    \end{itemize}
Therefore $\nabla_0 G_{00}$ is order $2$.

\item[3. $\nabla_a \nabla_b G_{00}$]
\beq\nn
\nabla_a \nabla_b G_{00} = \d_a(\nabla_b G_{00}) - \G^p_{ab} \nabla_p G_{00}-2\G^p_{a0}\nabla_b G_{p0}.
\eeq
Then,
\begin{itemize}
  \item By the above $\nabla_b G_{00}$ is order zero, therefore so is $\d_a(\nabla_b G_{00})$.
  \item The sum $\G^p_{ab} \nabla_p G_{00}$ contains terms with $p=i$ and $p=0$,
                                  \begin{itemize}
                                     \item When $p=i$ we have $\G^i_{ab}=\ordv{0}$ and $\nabla_i G_{00}=\ordv{0}$ so that their product is also order zero.
                                     \item When $p=0$ we have $\G^0_{ab}=\ordv{-2}$ (without using the derivative assumption) and $\nabla_0 G_{00}=\ordv{2}$ so that their product is order zero.
                                   \end{itemize}
        Thus $\G^p_{ab} \nabla_p G_{00}$ is order zero.
  \item $\G^p_{a0}$ is order zero, as is $\nabla_b G_{p0}$; therefore $\G^p_{a0}\nabla_b G_{p0}$ is order zero.
\end{itemize}
We can conclude that $\nabla_a \nabla_b G_{00}=\ordv{0}$.

\item[4. $\nabla_0 \nabla_0 G_{00}$]
\beq\nn
\nabla_0 \nabla_0 G_{00} = \d_0(\nabla_0 G_{00}) - \G^p_{00} \nabla_p G_{00}-2\G^p_{00}\nabla_b G_{00}.
\eeq
Then,
\begin{itemize}
  \item By the above $\nabla_0 G_{00}$ is order two, therefore so is $\d_0(\nabla_0 G_{00})$.
  \item The sum $\G^p_{00} \nabla_p G_{00}$ contains terms with $p=i$ and $p=0$,
                                  \begin{itemize}
                                     \item When $p=i$ we have $\G^i_{00}=\ordv{2}$ and $\nabla_i G_{00}=\ordv{0}$ so that their product is order two.
                                     \item When $p=0$ we have $\G^0_{00}=\ordv{0}$ and $\nabla_0 G_{00}=\ordv{2}$ so that their product is order two.
                                   \end{itemize}
        Thus $\G^p_{00} \nabla_p G_{00}$ is order two.
  \item The sum $\G^p_{00} \nabla_0 G_{p0}$ contains terms with $p=i$ and $p=0$,
                                  \begin{itemize}
                                     \item When $p=i$ we have $\G^i_{00}=\ordv{2}$ and $\nabla_0 G_{i0}=\ordv{0}$ so that their product is order two.
                                     \item When $p=0$ we have $\G^0_{00}=\ordv{0}$ and $\nabla_0 G_{00}=\ordv{2}$ so that their product is order two.
                                   \end{itemize}
        Thus $\G^p_{00} \nabla_0 G_{p0}$ is order two.
\end{itemize}
We can conclude that $\nabla_0 \nabla_0 G_{00}=\ordv{2}$.
\ed

We are now in a position to prove the theorem, which we do by taking the trace of the second derivative of $G_{00}$,
\bea
\triangle G_{00} &=& g^{ab}\nabla_a\nabla_b G_{00} \nn \\
&=& -\mv^{-2} \nabla_0\nabla_0 G_{00} + \sum_{i,j\in\{1,2,3\}}h_V^{ij}\nabla_i\nabla_j G_{00} \nn \\
&=& \ordv{0}. \nn
\eea
Then since $\v_B=\v_A/\v_A^{\overline{B}} + \orda{2}$, and since order is coordinate (and observer) independent we can conclude that,
$$\triangle G_{AB} = \ordc{0}.$$

\bibliography{Bib}

\begin{thebibliography}{10}

\bibitem{EinInfHoff:1938}
L.~Infeld A.~Einstein and B.~Hoffmann.
\newblock The gravitational equations and the problem of motion.
\newblock {\em Ann. Math.}, 39:65--100, 1938.

\bibitem{Cliff:1876}
W.~K. Clifford.
\newblock On the space theory of matter.
\newblock {\em Proc. Cambridge Phil. Soc.}, 2:157, 1876.
\newblock read in 1870.

\bibitem{radar3}
Luca~Lusanna David~Alba.
\newblock Generalized radar 4-coordinates and equal-time cauchy surfaces for
  arbitrary accelerated observers.
\newblock {\em Int.J.Mod.Phys.}, page 1149, 2007.

\bibitem{radar1}
Carl~E. Dolby and Stephen~F. Gull.
\newblock On radar time and the twin paradox.
\newblock {\em Am.J.Phys.}, 69:1257, 2001.

\bibitem{radar2}
Bahram~Mashhoon Donato~Bini, Luca~Lusanna.
\newblock Limitations of radar coordinates.
\newblock {\em Int.J.Mod.Phys.}, page 1413, 2005.

\bibitem{EinGrom:1927}
A.~Einstein and J.~Grommer.
\newblock motion of concentrations of mass-energy concluded not to be
  arbitrarily specifiable without violating field equations.
\newblock {\em Sitzber. Preuss. Akad. Wiss., Physik. Math}, K1:235, 1927.

\bibitem{ADM:1962}
S.~Deser R.~Arnowitt and C.W. Misner.
\newblock {\em Gravitation:an introduction to current research}, chapter~7,
  pages 227--265.
\newblock Wiley, 1962.

\bibitem{RiemG:1867}
B.~Riemann.
\newblock Ueber die hypothesen, welche der geometrie zu grunde liegen.
\newblock {\em Abh. Kgl. Ges. Wiss. Gött.}, 13, 1867.
\newblock Published pothumously by Richard Dedekind.

\bibitem{RiemE:1873}
B.~Riemann.
\newblock On the hypotheses which lie at the bases of geometry.
\newblock {\em Nature}, 8, 1873.
\newblock Translated by W. K. Clifford.

\bibitem{geons:1957}
John~A. Wheeler.
\newblock Geons.
\newblock {\em Phys. Rev.}, 97:511--536, 1955.

\bibitem{geometrodynamics:1961}
John~A. Wheeler.
\newblock Geometrodynamics and motion.
\newblock {\em Rev. Mod. Phys.}, 33:63--78, 1961.

\end{thebibliography}
\bibliographystyle{plain}

\end{document}